\documentclass[sigconf]{acmart}

\usepackage{acronym} 
\usepackage{algorithmic}
\usepackage{booktabs}
\usepackage{clrscode3e}
\usepackage{hyperref}
\usepackage{multirow}
\usepackage{graphicx}
\usepackage{subfigure}
\usepackage{textcomp}
\usepackage{xcolor}
\usepackage{tabularx}
\usepackage{xspace}
\usepackage{caption}
\usepackage[normalem]{ulem}
\usepackage{cleveref}
\usepackage{amsmath}
\usepackage{varwidth}
\usepackage{paralist}
\usepackage{wrapfig}
\usepackage{bm}
\usepackage[normalem]{ulem}
\usepackage{enumitem}
\usepackage{tikz}
\useunder{\uline}{\ul}{}
\usepackage[ruled, vlined, linesnumbered]{algorithm2e}
\usepackage{mathrsfs}
\usepackage{tabularx}
\usepackage{threeparttable}

\acrodef{NCDR}{Non-overlapping Cross-domain Recommendation}
\acrodef{MoE}{Mixture-of-Experts}
\acrodef{LLMs}{Large Language Models}
\acrodef{CDR}{Cross-Domain Recommendation}
\acrodef{PCME}{Preference-driven Conditioned Mixture-of-Experts}
\acrodef{RQ-VAE}{Residual-Quantized Variational Autoencoder}
\acrodef{DPQ}{Discrete Preference Quantizer}
\acrodef{PCR}{Preference-Conditioned Router}
\acrodef{DDM}{Domain Decoupling MoE}
\acrodef{MHA}{multi-head self-attention layer}
\acrodef{RQ}{Residual Quantization}
\acrodef{PQ}{Product Quantization}
\acrodef{DRM}{Decoupled Routing MoE}
\acrodef{LGCD}{Language-Guided Conditional Diffusion for CDR}
\acrodef{LPG}{LLM-based Pseudo-Interaction Generation}
\acrodef{CDPG}{Conditional Diffusion Preference Generator}
\AtBeginDocument{%
  \providecommand\BibTeX{{%
    \normalfont B\kern-0.5em{\scshape i\kern-0.25em b}\kern-0.8em\TeX}}}

\copyrightyear{2026}
\acmYear{2026}
\setcopyright{cc}
\setcctype{by}
\acmConference[SIGIR '26] {Proceedings of the 49th International ACM SIGIR Conference on Research and Development in Information Retrieval}{July 20--24, 2026}{Melbourne, VIC, Australia.}
\acmBooktitle{Proceedings of the 49th International ACM SIGIR Conference on Research and Development in Information Retrieval (SIGIR '26), July 20--24, 2026, Melbourne, VIC, Australia}
\acmISBN{979-8-4007-2599-9/2026/07}
\acmDOI{10.1145/XXXXXX.XXXXXX}

\begin{document}

\title[LGCD]{From Clues to Generation: Language-Guided Conditional Diffusion for Cross-Domain Recommendation}

\author{Ziang Lu}

\affiliation{
  \institution{Anhui University}
  \city{Hefei}
  \country{China}
}
\email{zianglu@outlook.com}

\author{Lei Sang}
\affiliation{
  \institution{Anhui University}
  \city{Hefei}
  \country{China}
}
\email{sanglei@ahu.edu.cn}

\author{Lin Mu}
\affiliation{
  \institution{Anhui University}
  \city{Hefei}
  \country{China}
}
\email{mulin@ahu.edu.cn}

\author{Yiwen Zhang}
\authornote{Corresponding Author.}
\affiliation{%
  \institution{Anhui University}
  \city{Hefei}
  \country{China}}
\email{zhangyiwen@ahu.edu.cn}

\renewcommand{\shortauthors}{Ziang Lu, Lei Sang, Lin Mu, and Yiwen Zhang}

\begin{abstract}
\ac{CDR} exploits multi-domain correlations to alleviate data sparsity. As a core task within this field, inter-domain recommendation focuses on predicting preferences for users who interact in a source domain but lack behavioral records in a target domain. Existing approaches predominantly rely on overlapping users as anchors for knowledge transfer. In real-world scenarios, overlapping users are often scarce, leaving the vast majority of users with only single-domain interactions. For these users, the absence of explicit alignment signals makes fine-grained preference transfer intrinsically difficult. To address this challenge, this paper proposes \ac{LGCD}, a novel framework that integrates \ac{LLMs} and diffusion models for inter-domain sequential recommendation. Specifically, we leverage LLM reasoning to bridge the domain gap by inferring potential target preferences for single-domain users and mapping them to real items, thereby constructing pseudo-overlapping data. We distinguish between real and pseudo-interaction pathways and introduce additional supervision constraints to mitigate the semantic noise brought by pseudo-interaction. Furthermore, we design a conditional diffusion architecture to precisely guide the generation of target user representations based on source-domain patterns. Extensive experiments demonstrate that LGCD significantly outperforms state-of-the-art methods in inter-domain recommendation tasks.
\end{abstract}



\maketitle

\section{Introduction}
Cross-domain recommendation addresses data sparsity and cold-start issues by leveraging user behaviors across correlated domains~\cite{chen2024survey, zang2022survey}. Unlike conventional intra-domain scenarios, inter-domain recommendation targets a more challenging setting: predicting preferences for users who are active in a source domain but have no interaction records in the target domain. Accurately recommending items for such cold-start users solely based on source-domain sequences remains a critical bottleneck in the field.


Existing approaches predominantly fall into two paradigms: \textbf{embedding and mapping methods} (e.g., EMCDR~\cite{man2017cross}, SSCDR~\cite{kang2019semi}) and \textbf{unified representation learning} (e.g., UniCDR~\cite{cao2023towards}, UCLR~\cite{yang2024not}). The former learns a projection function to map representations between domains, while the latter constructs a unified embedding space via contrastive learning or invariant feature extraction. However, both paradigms rely heavily on overlapping users as anchors for alignment. In real-world applications, overlapping users often constitute a sparse subset of the population, leaving the vast majority of users with only single-domain interactions. For these non-overlapping users, the absence of explicit alignment signals makes fine-grained preference transfer intrinsically difficult. While recent studies have attempted to utilize non-overlapping data~\cite{guo2023dan, guo2025automated, li2024mutual,lin2024mixed}, they primarily focus on mining general cross-domain knowledge rather than personalized transfer, often falling short in providing accurate recommendations for strictly cold-start users.

\begin{figure}[t]
\setlength{\abovecaptionskip}{0.1cm}
\setlength{\belowcaptionskip}{0.1cm} 
\centering
\includegraphics[width=8.5cm]{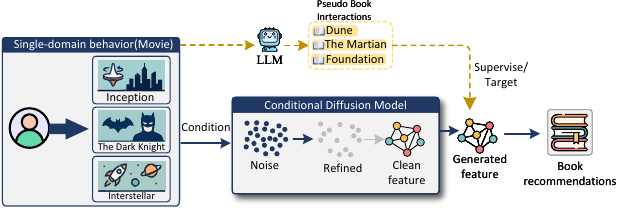}
\caption{The core idea of LGCD}
\label{fig:intro}
\end{figure}

To alleviate the dilemma of scarce overlapping anchors, we propose unlocking the potential of massive single-domain users as illustrated in Fig.~\ref{fig:intro}. Specifically, we leverage LLMs to infer potential cross-domain interests from source behaviors, creating "pseudo-overlapping" anchors to bridge the domain gap. Furthermore, inter-domain recommendation for cold-start users can be fundamentally modeled as a conditional generation task: creating accurate target user representations conditioned on source-domain patterns. To achieve this, we introduce Diffusion Models, which are renowned for their superior capabilities in complex distribution fitting and conditional generation~\cite{rombach2022high, croitoru2023diffusion, yang2024cross}. By treating source-domain features as conditions to guide the diffusion process, we can synthesize expressive latent user representations in the target domain, thereby enabling accurate preference prediction for cold-start users.

However, realizing this framework faces three robust challenges:

\textbf{(1) Misalignment between Open-ended Generation and Discrete Item Spaces.} LLMs output free-form text in an open-vocabulary space~\cite{wu2024survey, li2024large}, which does not naturally align with the discrete item set in the target domain. Without rigorous constraints, generated sequences may drift from the actual item space, rendering them unusable for recommendation.

\textbf{(2) Gap between Semantic and Collaborative Signals.} Pseudo-items generated by LLMs rely on textual semantics, often neglecting the latent collaborative signals (e.g., co-occurrence patterns) inherent in user behaviors~\cite{kim2025lost}. Naively treating these pseudo-interactions as ground truth introduces noise, which misguides the diffusion process, resulting in generated target preferences that deviate from the user's real intent.

\textbf{(3) Controllability of Diffusion-based Transfer.} While Diffusion Models are powerful, guiding them to generate user-specific target representations based on source-domain conditions is challenging. The model risks generating generic target-domain features that lose the personalized correlation with the source user, or conversely, strictly copying source patterns that do not fit the target domain's feature distribution.

To address these challenges, this paper proposes an \ac{LGCD} framework for inter-domain sequential recommendation.
\textbf{To tackle Challenge 1}, we encode the LLM-generated content into abstract preference features that share a unified semantic space with real item representations. We then employ semantic matching to map these continuous features onto the discrete target item set, thereby constructing valid pseudo-overlapping interactions.
\textbf{To address Challenge 2}, we prioritize real overlapping users as stable anchors to establish the correct cross-domain alignment direction. The LLM-generated pseudo-data acts as supplementary supervision, augmenting the model's ability to generate target preferences upon this foundational mapping. To effectively integrate them, we decouple their training pathways and employ a cyclic batch strategy, ensuring that the scarce real data consistently guides the distribution learning while the pseudo-data expands the preference coverage without introducing semantic noise.
\textbf{For Challenge 3}, we construct a conditional diffusion architecture centered on a cross-attention mechanism. This architecture incorporates source domain preference representations at each step of the denoising network to guide the target domain feature generation in a fine-grained manner, complemented by a \ac{MoE} fusion strategy to adaptively balance source-domain transferable preferences with target-domain specific patterns. Additionally, we train a lightweight guesser network to provide the semantic space prior of the target domain during the inference phase. This ensures that the generated target domain representations not only inherit abstract user preferences but also align with the target feature space, thereby enhancing CDR capabilities for cold-start users.

Our main contributions can be summarized as:
\begin{itemize}
\item We propose LGCD, a novel LLM- and diffusion-based framework for inter-domain sequential recommendation tailored to realistic single-domain user scenarios with scarce overlapping users.
\item We introduce an LLM-driven pseudo-overlap construction mechanism to synthesize anchors, complemented by a decoupled training strategy with cyclic batching. This approach effectively augments preferences using pseudo interactions while mitigating semantic noise by prioritizing real overlapping data for alignment.
\item We design a conditional diffusion architecture utilizing a cross-attention denoising network followed by a MoE fusion strategy. This structure precisely generates target domain preferences under source-domain guidance and adaptively integrates them to form the final user representation.
\item Extensive experiments on benchmark datasets demonstrate that LGCD significantly outperforms state-of-the-art methods in inter-domain recommendation tasks.
\end{itemize}
\section{Related Work}

\subsection{Cross-domain Recommendation}
Cross-domain recommendation is generally categorized into intra-domain and inter-domain settings. Intra-domain methods~\cite{li2024aiming, cao2022contrastive, li2020ddtcdr, zhu2019dtcdr, zhao2023cross} leverage auxiliary domain knowledge to enhance recommendations for users who are already active in the target domain. In contrast, inter-domain recommendation addresses a more challenging cold-start scenario where users lack any historical records in the target domain. This necessitates predicting target preferences relying exclusively on source-domain behaviors, presenting a significantly higher barrier for effective modeling compared to the intra-domain counterpart.

Existing work on inter-domain recommendation generally falls into two paradigms. The first is the mapping paradigm~\cite{man2017cross, kang2019semi, xuan2024diffusion, zhu2021transfer, zhu2022personalized, li2025disco, li2024cdrnp}, which utilizes overlapping users between domains as a bridge to learn mapping functions between user embeddings, thereby facilitating cross-domain information transfer. Among them, DisCo~\cite{li2025disco} projects the disentangled source-domain user intent embedding onto the target-domain through a decoder, and introduces intent-based contrastive learning to align the projected embedding with the user similarity structure of the target domain, effectively filtering out irrelevant source information. The second is the unified representation paradigm~\cite{cao2023towards, yang2024not, cao2022cross}, which leverages data from both domains to learn a unified representation, subsequently adapting these representations to specific domains to share common knowledge. For example, UCLR~\cite{yang2024not} employs a contrastive dual-stream collaborative autoencoder with user-aware contrastive learning and individualized temperatures to generate balanced user embeddings from pretrained global embeddings. 

However, the aforementioned methods rely heavily on overlapping users. Recent studies have begun to leverage non-overlapping users for cross-domain recommendation\cite{guo2023dan, guo2025automated, liu2024mcrpl, li2024mutual,lin2024mixed, wang2024making, chen2025leave}. For instance, PLCR~\cite{guo2025automated} introduces a domain-prompt mechanism between non-overlapping domains to learn domain-invariant knowledge, which is subsequently utilized to enhance performance in specific domains. Nevertheless, due to the scarcity of sufficient overlapping anchors, these approaches primarily reinforce the learning of similar knowledge across domains and struggle to achieve optimal performance in inter-domain recommendation scenarios.

\subsection{Diffusion Models in Recommendation}
Diffusion models exhibit superior capability in fitting complex data distributions by learning to reconstruct features from noise, achieving remarkable success in computer vision~\cite{ulhaq2022efficient, croitoru2023diffusion}. Recently, this paradigm has extended to sequential recommendation~\cite{li2023diffurec, liu2023diffusion, ma2024plug, wang2024conditional, yang2023generate}. For instance, DiffuRec~\cite{li2023diffurec} employs a diffusion process to generate distribution representations for historical items by corrupting the target item embedding, and a reverse process to reconstruct the target item representation for prediction, incorporating a rounding operation for mapping to discrete items. In parallel, existing works also apply diffusion models to cross-domain recommendation~\cite{xuan2024diffusion, li2025cd, li2025exploring, jin2025diffusion, li2025diffusion, liu2025llm}, where methods such as DMRec~\cite{li2025exploring} leverages a preference encoder to derive a guidance signal from a user's source domain interactions, which is then explicitly injected step-by-step into the diffusion model's reverse process to generate personalized user representations in the target domain. Nevertheless, these approaches invariably rely on overlapping users as anchors to guide the diffusion model in capturing the target domain distribution.

\begin{figure*}[t]
    \centering   
    \setlength{\abovecaptionskip}{0.1cm}
    \setlength{\belowcaptionskip}{0.1cm} 
    \includegraphics[width=\linewidth]{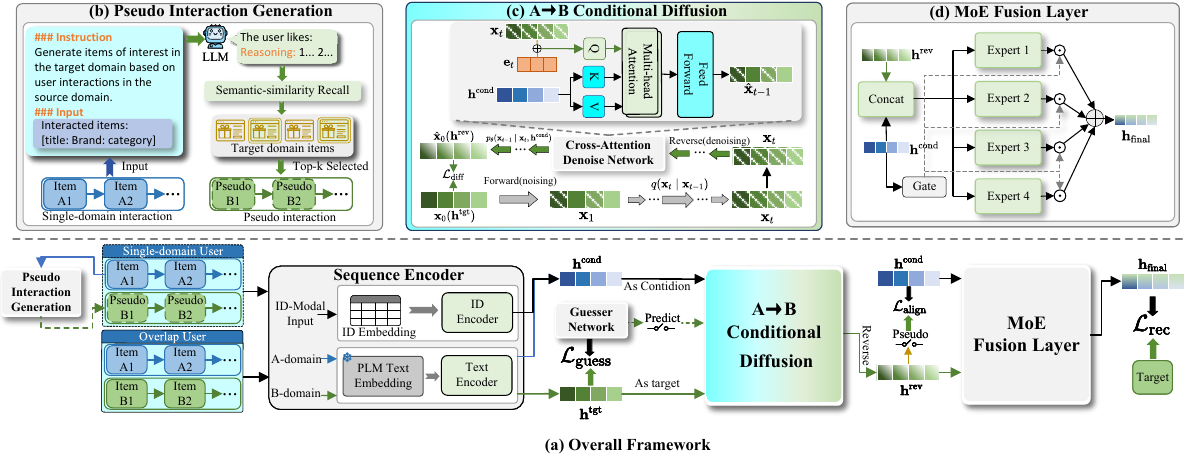}
    \caption{The overall framework of LGCD, illustrating the complete workflow with domain A as the source and domain B as the target. The ID-Modal Input comprises the complete interaction sequences from both domains for overlapping users, whereas for single-domain users, it consists solely of their observed source-domain interaction sequences. (b) The module synthesizes target-domain pseudo-interactions for single-domain users. (c) The module generates user preferences for the target domain under the supervision of conditional signals. (d) The module integrates conditional signals with the generated preferences to yield the final prediction.}
    \label{fig:overview}
\end{figure*}
\section{Methodology}
\subsection{Preliminaries}
Consider two distinct domains, A and B. These domains share a small subset of overlapping users, denoted as $\mathcal{U}^O$, while the majority of users interact exclusively within a single domain, represented as $\mathcal{U}^A$ and $\mathcal{U}^B$, respectively. Since domains A and B provide different services, their item sets, $\mathcal{V}^A$ and $\mathcal{V}^B$, are disjoint. For each item $v \in {\mathcal{V}^A}$ or ${\mathcal{V}^B}$, its associated textual information is represented as $c$. We denote the historical interaction sequence of a single-domain user $u_i^A \in \mathcal{U}^A$ as $S^A_i = \{v_1^A,v_2^A,\ldots,v_{m_i}^A\}$, and the sequence for an overlapping user $u^O_i \in \mathcal{U}^O$ as $S^O_i =  \{v_1^A, v_2^B, v_3^B, \ldots, v_{m_i}^A\}$. Our objective is to leverage the interaction data from both overlapping and non-overlapping users in the training set to learn a model. This model aims to recommend items in the target domain to users who have interactions solely in the source domain. Specifically, for a new user $u_j^A \in \mathcal{U}^A$, our goal is to predict and recommend items $v_i^B \in \mathcal{V}^B$.

\subsection{Overview of LGCD}
The framework of LGCD is shown in Fig.~\ref{fig:overview}, which comprises three core components: the \textit{LLM-based Pseudo-Interaction Generation Module}, the \textit{Conditional Diffusion Preference Generator} and the \textit{MoE Fusion Module}.

\textbf{1) \ac{LPG}.} This module leverages the reasoning capabilities of LLMs to bridge the data gap between domains. We prompt an LLM with the textual interaction history of a single-domain user to hypothesize potential interests within the target domain. To ground these open-ended textual predictions in the actual item space, we project them into a shared semantic space using a pre-trained text encoder and perform Top-k semantic retrieval against real target item text embeddings. The identified items are then integrated into the source sequence to construct a pseudo-cross-domain interaction trajectory.

\textbf{2) \ac{CDPG}.} This component functions as a conditional generative model, aiming to synthesize target-domain semantic preferences based on source-domain behaviors.
First, we construct a multi-modal conditional signal by fusing ID-based collaborative features and text-based semantic features from the source sequence.
This signal guides the diffusion process via a cross-attention-based denoising network, which iteratively injects source-domain preferences to ensure the generated target semantics remain consistent with the user's historical interests.
During training, we implement a decoupled strategy: for real overlapping users, we utilize ground-truth target semantics for direct supervision; for pseudo-overlapping users, we incorporate additional alignment constraints alongside the standard diffusion loss.
Additionally, a lightweight guesser network is trained to provide a target-space prior, stabilizing the inference process.

\textbf{3) MoE Fusion Module.} This module is designed to synthesize the final user representation for downstream recommendation. It employs the source-domain conditional representation to drive a gating network, which adaptively assigns weights to distinct experts. Each expert processes the concatenated features of the source condition and the diffusion-generated target representation. The final user embedding is obtained through a weighted aggregation of these expert outputs, effectively balancing transferable source knowledge with generated target-specific patterns.

\subsection{LLM-based Pseudo-Interaction Generation}
In real-world CDR scenarios, overlapping users are typically scarce, leaving the majority of users with interactions in only a single domain. The absence of explicit cross-domain anchors renders precise preference transfer for these single-domain users a significant challenge. To bridge this gap, we harness the semantic reasoning capabilities of LLMs, prompting them to infer potential interests in the target domain based on the textual semantics of users' historical interactions. However, given the stochastic nature and hallucination issues inherent in LLMs~\cite{huang2025survey, tonmoy2024comprehensive}, the generated content often fails to map directly to the discrete item set of the target domain. To address this mismatch, we employ a retrieval mechanism based on pre-trained text representations to align the generated preferences with existing target items, thereby constructing valid pseudo-overlapping interaction sequences.

\subsubsection{Pre-trained Sequence Encoder} \label{sec:seq_encoder}
We train independent sequential models for domains A and B using interaction sequences from users who are active solely within a single domain. Within the sequential model for each domain, we establish separate sequence encoders for the ID modality and the text modality, both of which adopt the Transformer~\cite{vaswani2017attention} architecture. Taking a single-domain user $u_i^A \in \mathcal{U}^A$ as an example,  we denote her ID-based interaction sequence as $S^A_i = \{v_1^A,v_2^A,\ldots,v_{m_i}^A\}$ and her text-based interaction sequence as $C^A_i = \{c_1^A,c_2^A,\ldots,c_{m_i}^A\}$, where $c^A$ represents the textual information of $v^A$, including its title, category, and brand. The collaborative sequence representation and the semantic sequence representation derived from these two interaction modalities are formulated as follows:
\begin{gather}
\mathbf{h}_i^{\text{ID}} = \text{Encoder}_{\text{ID}}^A(S_i^A) ,\quad
\mathbf{h}_i^{\text{text}} = \text{Encoder}_{\text{text}}^A(C_i^A)  \\
\mathbf{h}_i^{\text{fusion}} = \mathbf{W}_{\text{fusion}}^A[\mathbf{h}_i^{\text{ID}} \oplus \mathbf{h}_i^{\text{text}}]
+ \mathbf{b}_{\text{fusion}}^A. \label{eq:fusion_layer}
\end{gather}
where $\mathbf{h}_i^{\text{ID}}, \mathbf{h}_i^{\text{text}} \in \mathbb{R}^d$ are sequence representation of two modalities, $\mathbf{W}^A_\text{fusion} \in \mathbb{R}^{d \times 2d}$, $\mathbf{b}^A_\text{fusion} \in \mathbb{R}^d$ are learnable parameters. The sequence representation after the fusion of two modalities is $\mathbf{h}_i^\text{fusion}$, which we use for loss calculation:
\begin{equation}
    \mathcal{L}_{\text{pre}}^A=-\frac{1}{\left|\mathcal{S}_{A}\right|} \sum_{S_{i}^{A} \in \mathcal{S}_{A}} \log P\left(v_{i+1}^A \mid S_{i}^{A}\right),
\end{equation}
where $\left|\mathcal{S}_{A}\right|$ is the number of single domain sequences in domain A, $P\left(v_{i+1}^A \mid S_{i}^{A}\right) = \text{Softmax} (\mathbf{h}_i^\text{fusion} \cdot \mathbf{E}^A_\text{fusion})$, $\mathbf{E}^A_\text{fusion} = \mathbf{W}_\text{fusion} [\mathbf{E}^A_\text{ID} \oplus \mathbf{E}^A_\text{text}] + \mathbf{b}_\text{fusion}$ is the fusion embedding of all items in domain A. We pre-train sequence models in the A and B domains using $\mathcal{L}_{\text{pre}}^A$ and $\mathcal{L}_{\text{pre}}^B$, respectively.

\subsubsection{Pseudo Interaction Generation by LLM}
To leverage the textual reasoning capabilities of LLMs for pseudo-item generation, we construct instruction prompts and feed the user's textual interaction sequence into the LLM. This prompts the model to generate several pseudo-textual interactions for single-domain users. For a specific user $u_i^A \in \mathcal{U}^A$, this process is formulated as:
\begin{equation}
    \widetilde{C}_i^B = \{\tilde{c}_1^B, \tilde{c}_2^B, \ldots, \tilde{c}_{m_g}^B\} = \operatorname{LLM}(C_i^A),
\end{equation}
where $\tilde{c}_j^B$ is the text of an item generated by LLM, these generated texts are the outcomes of LLM-based textual reasoning and may deviate from actual items. To facilitate matching within the real item collection, it is necessary to unify the pseudo-text sequences and the textual features of the target domain into a continuous semantic space. We input $\widetilde{C}_i^B$ into the pre-trained text encoder to obtain the textual sequence representation for user $u_i^A$ in the B domain:
\begin{equation}
    \tilde{\mathbf{h}}_i^B = \text{Encoder}^B_\text{text}(\widetilde{C}_i^B).
\end{equation}
Subsequently, based on cosine similarity, we retrieve the $n_K$ most similar real-world items from the B domain's text embedding pool:
\begin{equation}
    S_i^B = \{v_1^B, v_2^B, \ldots, v_{n_K}^B\} = \operatorname{Top-k} \{\text{cos}\langle\tilde{\mathbf{h}}_i^B , \mathbf{e}_j^B \rangle \mid  \mathbf{e}_j^B \in \mathbf{E}^B_\text{text} \}.
\end{equation}
Finally, we randomly insert the recalled pseudo interaction items into the original sequence $S_i^A$ to obtain the pseudo overlapping interaction sequence $\tilde{S}_i^O = \{v_1^A, v_1^B, v_2^B, v_2^A, \ldots\}$ of user $u_i^A$.

\subsection{Conditional Diffusion Preference Generator}
With the pseudo-overlapping interactions constructed, our goal is to train a conditional diffusion model that synthesizes target-domain semantic preferences conditioned on multi-modal source signals. However, simply treating noisy pseudo-interactions as ground truth risks biasing the generative distribution learned by the model. To mitigate this, we devise a dual-path training strategy that differentiates between real and pseudo-overlapping users. Specifically, within the pseudo-data pathway, we mask the ID modality of the generated items and incorporate an auxiliary alignment regularization loss. This ensures that the model leverages the augmented data for preference transfer without misguiding the generation process into noisy or invalid semantic regions.
\subsubsection{Construction of Conditional Signals and Target Features}
We determine the construction of target features based on the domain to which the last item in the interaction sequence belongs. Specifically, we utilize the user's interactions within that specific domain to formulate the target representation. For instance, consider an overlapping user $u_j^O$ with an interaction sequence $S_j^O = \{v_1^A, v_2^B, \ldots, v_{m_j}^B\}$. The final interaction item $v_{m_j}^B \in \mathcal{V}^B$, we extract the user's textual interaction sequence $C_j^B = \{c_2^B, c_3^B, \ldots\}$ specifically from domain B. This sequence $C_j^A$ is then input into an encoder to generate the textual sequence representation, which serves as the target feature.
\begin{equation}
    \mathbf{h}_j^\text{tgt} = \text{Encoder}_\text{text}^B(C_j^B).
\end{equation}
The construction of conditional signals integrates both collaborative and semantic information derived from user historical interactions. For user $u_j^O$, we first input her complete interaction $S_j^O$ into $\text{Encoder}_\text{ID}^A$ to obtain the collaborative sequence representation $\mathbf{h}_j^\text{ID}$. Subsequently, we extract the A domain interactions from $C_j^O$ to form $C_j^A$, which is fed into $\text{Encoder}_\text{text}^A$ to generate the semantic sequence representation (The structures of the encoders mentioned above are the same as those in Section~\ref {sec:seq_encoder}). We combine these two sequence representations as conditional signals for the diffusion model:
\begin{equation}
    \mathbf{h}_j^\text{cond} = f_\text{fusion}^A(\mathbf{h}_j^\text{ID} \oplus \mathbf{h}_j^\text{text}),
\end{equation}
where $f_\text{fusion}^A$ adopts the same structure as Eq.~(\ref{eq:fusion_layer}). 

For pseudo-overlapping interaction sequences, the distinction lies in the acquisition of the collaborative sequence representation. Considering that the pseudo-items are derived via semantic reasoning and inserted randomly into the sequence, they lack valid temporal information. Consequently, incorporating them into the modeling of the collaborative sequence representation would deviate from the user's original collaborative preferences. Therefore, we exclusively utilize the user's authentic interactions from the source domain to construct the conditional signal.

\subsubsection{Forward Process}
We adopt one-dimensional scalar noise scheduling $\{\beta_t\}_{t=1}^T$, in which the linearly increasing $\beta_t$ is obtained by uniformly interpolating and squaring over the $[\sqrt{\beta_{\min}},\sqrt{\beta_{\max}}]$ interval.
Given the target feature $\mathbf{x}_0 = \mathbf{h}^\text{tgt}$, the forward diffusion process is defined as a Markov chain, where the single-step transition is formulated as:
\begin{equation}
    q(\mathbf{x}_t \mid \mathbf{x}_{t-1})
= \mathcal{N}\!\left(\sqrt{\alpha_t}\,\mathbf{x}_{t-1},\,\beta_t\mathbf{I}\right),
\quad \alpha_t = 1-\beta_t.
\end{equation}
Since the Gaussian distribution remains closed under linear transformations and additive Gaussian noise, the aforementioned equation yields a closed-form marginal distribution from $\mathbf{x}_0$ to any arbitrary time step $\mathbf{x}_t$:
\begin{equation}
    q(\mathbf{x}_t \mid \mathbf{x}_0)
= \mathcal{N}\!\left(\sqrt{\bar{\alpha}_t}\,\mathbf{x}_0,\,(1-\bar{\alpha}_t)\mathbf{I}\right),
\quad \bar{\alpha}_t=\prod_{\tau=1}^{t}\alpha_\tau .
\end{equation}
The noise state $\mathbf{x}_t$ generated at any time step $t$ from $\mathbf{x}_0$ is:
\begin{equation}
    \mathbf{x}_t=\sqrt{\bar{\alpha}_t}\mathbf{x}_0+\sqrt{1-\bar{\alpha}_t}\boldsymbol{\epsilon},\quad\boldsymbol{\epsilon}\sim\mathcal{N}(\mathbf{0},\mathbf{I}).
\end{equation}
This process progressively corrupts the semantic features of the target domain. Consequently, the data approximates Gaussian noise at large time steps while preserving partial structural information at small time steps. This mechanism yields training samples of varying difficulty for the subsequent conditional denoising phase.

\subsubsection{Cross-Attention Denoising Network}
Motivated by the manifold assumption~\cite{li2025back}, we configure the denoising network to directly regress target features rather than estimating noise. This approach concentrates the model's capacity on the user preference manifold, thereby mitigating the curse of dimensionality inherent in high-dimensional sequence modeling. Furthermore, it ensures that the generated target-domain semantic representations naturally reside within a representation subspace that aligns with ground-truth user preferences. We demonstrate the superiority of this strategy in Section~\ref{sec:diff_tgt}.

In the denoising network, we first embed the time step and inject it into the noisy features:
\begin{equation}
    \tilde{\mathbf{x}}_t = \mathbf{x}_t + \mathbf{e}_t.
\end{equation}
Then utilized a cross-attention network to guide denoising generation with conditional signals. Specifically, we let $\tilde{\mathbf{x}}_t$ as the query and conditional signal $\mathbf{h}^\text{cond}$ as the key and value:
\begin{equation}
    \hat{\mathbf{x}}_t = \text{CrossAttn}( \tilde{\mathbf{x}}_t, \mathbf{h}^\text{cond}, \mathbf{h}^\text{cond}).
\end{equation}
Subsequently, residual connections are used to obtain the output of the first stage:
\begin{equation}
    \mathbf{x}_t^{(1)}=\tilde{\mathbf{x}}_t+\hat{\mathbf{x}}_t.
\end{equation}
We use the feed-forward network and residual connections to obtain the final output of the denoising network:
\begin{equation}
    \mathbf{x}_t^{(2)} = \mathbf{x}_t^{(1)} + \text{FFN}(\mathbf{x}_t^{(1)}).
\end{equation}
The estimation of clean target semantic features by the denoising network during this entire process can be expressed as:
\begin{equation}
    \hat{\mathbf{x}}_0=f_\theta(\mathbf{x}_t,t,\mathbf{Z}^\mathrm{cond})=\mathbf{x}_t^{(2)}
\end{equation}
During the training phase, we make the denoising network directly regress to the clean feature $\mathbf{x}_0 = \mathbf{h}^\text{tgt}$, that is, optimize the following diffusion loss:
\begin{equation}
\mathcal{L}_{\mathrm{diff}}=\mathbb{E}_{\mathbf{x}_0,t,\boldsymbol{\epsilon}}\left[\left\|f_\theta\left(\sqrt{\bar{\alpha}_t}\mathbf{x}_0+\sqrt{1-\bar{\alpha}_t}\mathbf{\epsilon},t,\mathbf{h}^{\mathrm{cond}}\right)-\mathbf{x}_0\right\|_2^2\right],
\end{equation}

\subsubsection{Reverse Process}
In the reverse process, we start with an initial representation $\mathbf{h}^\text{init}$ (instantiated as $\hat{\mathbf{x}}_0$ during training and synthesized by the guesser network introduced in Section~\ref{sec:guess} during testing). 
Given a total of $T$ diffusion steps, we introduce a start-step ratio $\lambda \in (0, 1]$ and determine the corresponding starting time step as $t_\text{start} = \lambda T$.
We first forward-diffuse $\mathbf{h}^\text{init}$ to this intermediate time step $t_\text{start}$, obtaining a partially noised representation $\mathbf{x}_{t_\text{start}}$. 

Subsequently, at each step $t = \{t_\text{start}, \ldots, 0\}$, we construct an approximate posterior distribution by leveraging the output from the denoising network alongside pre-computed diffusion parameters:
\begin{gather}
    p_\theta(\mathbf{x}_{t-1}\mid\mathbf{x}_t,\mathbf{h}^\mathrm{cond})
    = \mathcal{N}\!\big(\boldsymbol{\mu}_\theta(\mathbf{x}_t,t,\mathbf{h}^\mathrm{cond}),\boldsymbol{\Sigma}_t\big), \\
    \boldsymbol{\mu}_\theta(\mathbf{x}_t,t,\mathbf{h}^{\mathrm{cond}})=\frac{\sqrt{\alpha_{t-1}}\left.\beta_t\right.}{1-\bar{\alpha}_t}\hat{\mathbf{x}}_0+\frac{\sqrt{\alpha_t}\left(1-\bar{\alpha}_{t-1}\right)}{1-\bar{\alpha}_t}\mathbf{x}_t
\end{gather}
After a complete reverse process, we obtain the target feature $\mathbf{h}^\text{rev}$ restored by the diffusion model. 
By controlling $t_\text{start}$ through $\lambda$, we ensure that the macroscopic semantic contours are preserved even after noise injection. This strategy enables the diffusion module to perform stable and effective cross-domain semantic refinement upon the coarse target domain representation, effectively mitigating the instability inherent in reconstructing features from pure noise.

For pseudo overlapping users, we also add an additional constraint loss at this point:
\begin{equation}
    \mathcal{L}_{\mathrm{align}}
    = \left\|\mathbf{h}^{\mathrm{rev}}-\mathbf{h}^{\mathrm{cond}}\right\|_2^2,
\end{equation}
This constraint encourages the generated target-domain semantics to maintain cross-domain style transformation without deviating excessively from the conditional representations determined by real source-domain interactions. 
\subsubsection{Guesser network} \label{sec:guess}
To provide a more plausible initialization for the reverse diffusion process during the inference phase, we introduce two lightweight linear guessing networks, denoted as $g_{A \to B}$ and $g_{B \to A}$, concurrently with the diffusion training. These networks function to learn the mapping from conditional signals to the semantic space of the target domain. During diffusion training, for the path $A \to B$, the training loss for the guessing networks is formulated as:
\begin{equation}
    \mathcal{L}_{\mathrm{guess}}=\left\|g_{A\to B}(\mathbf{h}^{\mathrm{cond}})-\mathbf{h}^{\mathrm{tgt}}\right\|_2^2,
\end{equation}
with the $B \to A$ path following an analogous formulation. During the inference phase, as the ground-truth target domain text sequence is inaccessible, we utilize the output of the guessing network as the initial representation for the reverse diffusion process:
\begin{equation}
    \mathbf{h}_{\mathrm{init}} =
    \begin{cases}
        g_{A\to B}(\mathbf{h}^{\mathrm{cond}}), & \mathrm{A\to B},\\
        g_{B\to A}(\mathbf{h}^{\mathrm{cond}}), & \mathrm{B\to A}.
    \end{cases}
\end{equation}
This strategy not only provides a reasonable prior of the target domain space for the diffusion model but also reduces the required number of reverse steps and alleviates training difficulty.

\subsection{MoE Fusion Module}
Upon obtaining the target-domain representation via the diffusion model, we employ a MoE module to integrate it with the source-domain conditional signal. Our design acknowledges that the conditional signal derived from actual interactions faithfully captures stable user preference patterns. Conversely, the generated representation inevitably carries residual noise and may deviate from the true distribution in local feature spaces. Therefore, we strategically utilize only the robust conditional signal to drive the gating network. This strategy ensures that the expert selection mechanism remains anchored in the user's deterministic behaviors and prevents the fusion process from being misled by generative features. Taking user $u_i^A$ as an example, we set $n_e$ experts with the following weights:
\begin{equation}
    \alpha_i = [\alpha_{i,1}, \ldots, \alpha_{i,n_e}]= \mathrm{Softmax}(\mathbf{W}_\text{gate} \mathbf{h}_i^\text{cond}),
\end{equation}
where $\mathbf{W}_\text{gate} \in \mathbb{R}^{d \times 1}$. We concatenate the generated features $\mathbf{h}_i^\text{cond}$ with the conditional signal to serve as the shared input for the experts. This configuration enables each expert, under the supervision of the condition, to extract and filter salient information from the generated features, followed by a weighted aggregation process.
\begin{equation}
    \mathbf{h}_i^\text{final}=\sum_{n=1}^{n_e}\alpha_{i,n}\text{Expert}([\mathbf{h}_i^\text{cond} \oplus \mathbf{h}_i^\text{rev}]) \in \mathbb{R}^d,
\end{equation}
where each expert is structured as a linear mapping network projecting from $2d$ to $d$ dimensions. We utilize the output of the MoE module $ \mathbf{h}_i^\text{final}$, as the final representation to compute the task loss:
\begin{equation}
    \mathcal{L}_{\text{rec}}=-\frac{1}{\left|\mathcal{S}_{A}\right|} \sum_{S_{i}^{A} \in \mathcal{S}_{A}} \log P\left(v_{i+1}^B \mid S_{i}^{A}\right),
\end{equation}
where $P\left(v_{i+1}^B \mid S_{i}^{A}\right) = \text{Softmax}(\mathbf{h}_i^\text{final} \cdot \mathbf{E}^B_\text{fusion})$. We train the model by jointly optimizing multiple losses.For the real overlapping user set $\mathcal{U}^O$, the total loss is:
\begin{equation}
    \mathcal{L}_\text{total}^O = \mathcal{L}_\text{diff} + \mathcal{L}_\text{guess} + \mathcal{L}_\text{rec}. 
\end{equation}
For the set of single domain interactive users $\mathcal{U}^A$ and  $\mathcal{U}^B$, the total can be:
\begin{equation}
    \mathcal{L}_\text{total}^{A \cup B} = \mathcal{L}_\text{diff} + \mathcal{L}_\text{guess} +  \mathcal{L}_\text{align} + \mathcal{L}_\text{rec}. 
\end{equation}
Acknowledging that real overlapping users constitute a minor portion of the overall population yet serve as the primary source of supervision for learning cross-domain mappings, we employ a cyclic overlapping sampling strategy during mini-batch construction. Specifically, we enforce the inclusion of a fixed number of real overlapping instances in each training batch. This strategy inherently performs an implicit re-weighting of the loss for overlapping samples, ensuring that the model consistently receives gradient signals for cross-domain alignment within limited training iterations.
\section{Experiments}
\begin{table*}[]
\setlength\tabcolsep{2.5pt}
\setlength{\belowcaptionskip}{0.1cm} 
\centering
\caption{Overall performance comparison on Movie-Book and Food-Kitchen datasets. The best results are indicated in bold, and the second-best results are underlined. Improvements over baselines are statistically significant with $p < 0.001$.}
\label{tab:compare}
\vspace{-3mm}
\resizebox{\textwidth}{!}{
\begin{tabular}{l|cccc|cccc|cccc|cccc}
\hline
\multicolumn{1}{l|}{\multirow{3}{*}{Method}} &
  \multicolumn{4}{c}{\textbf{Movie}} &
  \multicolumn{4}{c|}{\textbf{Book}} &
  \multicolumn{4}{c}{\textbf{Food}} &
  \multicolumn{4}{c}{\textbf{Kitchen}} \\ \cline{2-17} 
\multicolumn{1}{c|}{} &
  \multicolumn{2}{c}{HR} &
  \multicolumn{2}{c|}{NDCG} &
  \multicolumn{2}{c}{HR} &
  \multicolumn{2}{c|}{NDCG} &
  \multicolumn{2}{c}{HR} &
  \multicolumn{2}{c|}{NDCG} &
  \multicolumn{2}{c}{HR} &
  \multicolumn{2}{c}{NDCG} \\\cline{2-17} 
\multicolumn{1}{c|}{} &
  H@5 &
  H@10 &
  N@5 &
  N@10 &
  H@5 &
  H@10 &
  N@5 &
  N@10 &
  \multicolumn{1}{c}{H@5} &
  \multicolumn{1}{c}{H@10} &
  \multicolumn{1}{c}{N@5} &
  \multicolumn{1}{c|}{N@10} &
  \multicolumn{1}{c}{H@5} &
  \multicolumn{1}{c}{H@10} &
  \multicolumn{1}{c}{N@5} &
  \multicolumn{1}{c}{N@10} \\ \hline 
GRU4Rec &
  0.0161 &
  0.0354 &
  0.0085 &
  0.0147 &
  0.0286 &
  0.0533 &
  0.0169 &
  0.0248 &
  0.0367 &
  0.0867 &
  0.0236 &
  0.0398 &
  0.0691 &
  0.1146 &
  0.0438 &
  0.0582 \\
DiffuRec &
  {\ul 0.0357} &
  0.0450 &
  {\ul 0.0266} &
  0.0297 &
  0.0057 &
  0.0152 &
  0.0031 &
  0.0061 &
  0.0600 &
  0.1267 &
  0.0313 &
  0.0520 &
  0.0392 &
  0.0612 &
  0.0226 &
  0.0300  \\
SASRec &
  0.0354 &
  0.0579 &
  0.0218 &
  0.0291 &
  0.0248 &
  0.0438 &
  0.0161 &
  0.0221 &
  0.1900 &
  0.2900 &
  0.1187 &
  0.1509 &
  0.2276 &
  0.3250 &
  0.1522 &
  0.1836  \\
FEARec &
  0.0289 &
  0.0482 &
  0.0169 &
  0.0227 &
  0.0289 &
  0.0482 &
  0.0169 &
  0.0227 &
  0.1733 &
  0.2733 &
  0.1044 &
  0.1363 &
  0.2402 &
  0.3454 &
  0.1530 &
  0.1866 \\ 
UniSRec &
  0.0161 &
  0.0386 &
  0.0091 &
  0.0161 &
  0.0305 &
  0.0457 &
  0.0218 &
  0.0264 &
  0.0733 &
  0.1235 &
  0.0444 &
  0.0596 &
  0.0675 &
  0.1020 &
  0.0451 &
  0.0562 \\ \hline
Tri-CDR &
  0.0225 &
  0.0354 &
  0.0141 &
  0.0182 &
  0.0210 &
  0.0476 &
  0.0123 &
  0.0210 &
  0.1733 &
  0.2967 &
  0.1094 &
  0.1483 &
  0.1538 &
  0.2449 &
  0.1077 &
  0.1296 \\ 
MGCL &
  0.0129 &
  0.0289 &
  0.0089 &
  0.0140 &
  0.0229 &
  0.0381 &
  0.0150 &
  0.0199 &
  0.2467 &
  0.3367 &
  0.1673 &
  0.1964 &
  {\ul 0.2590} &
  0.3140 &
  {\ul 0.1910} &
  {\ul 0.2085} \\
CTT &
  0.0322 &
  0.0579 &
  0.0220 &
  0.0302 &
  0.0362 &
  {\ul 0.0838} &
  0.0237 &
  0.0393 &
  0.1567 &
  0.2400 &
  0.0969 &
  0.1237 &
  0.1567 &
  0.2400 &
  0.0969 &
  0.1237 \\ \hline
DA-DAN &
  0.0289 &
  0.0643 &
  0.0193 &
 {\ul 0.0309} &
  {\ul 0.0438} &
  0.0724 &
 {\ul 0.0318} &
  {\ul 0.0408} &
  0.1633 &
  0.2733 &
  0.1124 &
  0.1475 &
  0.2480 &
  {\ul 0.3705} &
  0.1648 &
  0.2037 \\ 
LLMCDSR &
  0.0257 &
  0.0418 &
  0.0207 &
  0.0259 &
  0.0286 &
  0.0438 &
  0.0165 &
  0.0215 &
  0.2200 &
  0.3600 &
  0.1443 &
  0.1889 &
  0.2214 &
  0.3155 &
  0.1541 &
  0.1847 \\
PLCR &
  0.0322 &
  0.0514 &
  0.0202 &
  0.0264 &
  0.0381 &
  0.0571 &
  0.0271 &
  0.0331 &
  {\ul 0.2600} &
  {\ul 0.3967} &
  {\ul 0.1647} &
  {\ul 0.2085} &
  0.2308 &
  0.3375 &
  0.1540 &
  0.1883 \\ \hline
SSCDR &
  0.0233 &
  0.0397 &
  0.0179 &
  0.0232 &
  0.0338 &
  0.0565 &
  0.0212 &
  0.0303 &
  0.1943 &
  0.2747 &
  0.1264 &
  0.1455 &
  0.1941 &
  0.2827 &
  0.1465 &
  0.1709 \\
UniCDR &
  0.0330 &
  0.0511 &
  0.0197 &
  0.0265 &
  0.0371 &
  0.0621 &
  0.0257 &
  0.0377 &
  0.2328 &
  0.3109 &
  0.1505 &
  0.1785 &
  0.2228 &
  0.3222 &
  0.1689 &
  0.1903 \\
UCLR &
  0.0342 &
  0.0509 &
  0.0193 &
  0.0270 &
  0.0353 &
  0.0632 &
  0.0255 &
  0.0387 &
  0.2288 &
  0.3310 &
  0.1567 &
  0.1916 &
  0.2213 &
  0.3323 &
  0.1677 &
  0.1920 \\
CD-CDR &
  0.0361 &
 {\ul 0.0675} &
  0.0225 &
  0.0293 &
  0.0412 &
  0.0681 &
  0.0305 &
  0.0400 &
  0.2467 &
  0.3233 &
  0.1565 &
  0.1816 &
  0.2559 &
  0.3250 &
  0.1817 &
  0.1977 \\ \hline
\textbf{LGCD} &
  \textbf{0.0482} &
  \textbf{0.0772} &
  \textbf{0.0286} &
  \textbf{0.0381} &
  \textbf{0.0590} &
  \textbf{0.0914} &
  \textbf{0.0373} &
  \textbf{0.0477} &
  \textbf{0.2800} &
  \textbf{0.4000} &
  \textbf{0.1805} &
  \textbf{0.2190} &
  \textbf{0.3093} &
  \textbf{0.3893} &
  \textbf{0.2140} &
  \textbf{0.2397} \\
\textit{Improv.} &
  \textit{+35.0}\% &
  \textit{+14.4}\% &
  \textit{+7.52}\% &
  \textit{+23.3}\% &
  \textit{+34.7}\% &
 \textit{+9.07}\% &
 \textit{+17.3}\% &
  \textit{+16.9}\% &
  \textit{+7.69}\% &
  \textit{+0.83}\% &
  \textit{+9.59}\% &
  \textit{+5.04}\% &
  \textit{+19.4}\% &
  \textit{+5.07}\% &
  \textit{+12.0}\% &
  \textit{+15.0}\% \\ \hline
\end{tabular}
}
\label{main-exp}
\end{table*}
\subsection{Experimental Setup}
\subsubsection{Datasets and Evaluation Protocol}



\begin{table}[t]
\setlength{\abovecaptionskip}{0.1cm}
  \setlength{\belowcaptionskip}{0.1cm} 
\centering

\setlength{\tabcolsep}{3pt}
\setlength{\arrayrulewidth}{0.4pt} 
\caption{Statistics of the two cross-domain datasets.}
\label{tab:data-stat}
\begin{tabular}{l|cc|cc}
\hline
\multirow{2}{*}{Statistic} &
\multicolumn{2}{c|}{Movie-Book} &
\multicolumn{2}{c}{Food-Kitchen} \\
\cline{2-5}
& Movie & Book & Food & Kitchen \\
\hline
\# Users (non-overlap) & 6{,}819  & 50{,}242 & 5{,}041 & 35{,}889 \\
\# Users (overlap)     & \multicolumn{2}{c|}{2{,}666} & \multicolumn{2}{c}{5{,}781} \\
\hline
\# Interactions (single) & 12{,}185 & 124{,}889 & 5{,}472 & 38{,}191 \\
\# Interactions (cross)  & \multicolumn{2}{c|}{6{,}652}  & \multicolumn{2}{c}{7{,}501} \\
\hline
\# Items             & 12{,}875 & 93{,}860 & 8{,}661 & 27{,}637 \\
Avg.\ $m$            & \multicolumn{2}{c|}{9.01} & \multicolumn{2}{c}{8.22} \\
\hline
\end{tabular}
\end{table}
We construct two cross-domain datasets, i.e., \textbf{Movie-book} and \textbf{Food-Kitchen}, derived from the public Amazon\footnote{https://nijianmo.github.io/amazon/index.html} datasets. Following the previous work suggest~\cite {xin2025llmcdsr}, for the Movie-Book pair, we utilize interaction data from the most recent month and apply a 10-core filtering strategy to remove users and items with fewer than ten interactions. For the Food-Kitchen pair, we employ data spanning the past year and apply a 5-core filter. For these users, we identify the domain of the last interacted item and filter all prior interactions within that specific domain. This ensures that the user's historical interaction sequence and the ground-truth item reside in different domains. The remaining 80\% of overlapping interaction sequences, combined with all single-domain interaction sequences, constitute the training set. Table 2 summarizes the statistics of the datasets used in our experiments.

In the evaluation phase, we randomly sample 999 negative items from all participating domains to form a candidate list alongside the ground-truth item. We employ HR@N and NDCG@N as evaluation metrics, reporting results for $\text{N} \in \{5, 10\}$.

\subsubsection{Baseline Methods}
We compared LGCD with the following three types of baselines.
\begin{itemize}[leftmargin=*]
\item \textbf{Single-domain methods}: GRU4Rec~\cite{hidasi2015session}, DiffuRec~\cite{li2023diffurec}, SASRec~\cite{kang2018self}, FEARec~\cite{du2023frequency}, UniSRec~\cite{hou2022towards}.
\item \textbf{Intra-domain methods}: Tri-CDR~\cite{ma2024triple}, MGCL~\cite{xu2025multi}, CTT~\cite{zhou2025contrastive}.
\item \textbf{Non-overlapping methods}: DA-DAN~\cite{guo2023dan}, LLMCDSR~\cite{xin2025llmcdsr}, PLCR~\cite{guo2025automated}.
\item \textbf{Inter-domain methods}: SSCDR~\cite{kang2019semi}, UniCDR~\cite{cao2023towards}, UCLR~\cite{yang2024not}, CD-CDR~\cite{li2025cd}.
\end{itemize}
To ensure a fair comparison, we train all baselines, including single-domain recommendation models, on the identical dataset employed by LGCD and evaluate them under the same cross-domain task settings.


\subsubsection{Implementation Details}
We implement LGCD using the PyTorch framework and conduct all experiments on an NVIDIA RTX 4090 GPU. We set the hidden dimension size $d$ to 64. The model is optimized using the Adam~\cite{kingma2014adam} optimizer with a learning rate of 0.001. Regarding data sampling, we sample single-domain interactions with a batch size of 128. For overlapping interactions, we maintain a separate iterator with a batch size of 64. This iterator will repeatedly load limited data, ensuring the continuous participation of overlapping data throughout the training process. In the LPG module, we employ Baichuan2-7B-Chat\footnote{https://huggingface.co/baichuan-inc/Baichuan2-7B-Chat} as the inference LLM and utilize jina-embeddings-v2-base-en\footnote{https://huggingface.co/jinaai/jina-embeddings-v2-base-en} to encode the generated text. We select the $n_K = 10$ most similar items to serve as pseudo-items. Within the CDPG module, the noise schedule parameters $\beta_{min}$ and $\beta_{max}$ are configured to 0.001 and 0.1, respectively. The denoising network employs 4 attention heads, and the number of reverse generation steps $T$ is set to 100. The hyperparameter $\lambda$ is set to 0.7 for the Movie and Book domains, 0.5 for the Food domain, and 0.3 for the Kitchen domain. For the MoE Fusion module, we designate the number of experts as 8. 


\begin{table*}
  \centering
  \setlength{\abovecaptionskip}{0.1cm}
  \setlength{\belowcaptionskip}{0.1cm} 
  \caption{Ablation studies of LGCD on Movie-Book and Food-Kitchen datasets.}
  \label{tab:ablation}
  
  \newcolumntype{Y}{>{\centering\arraybackslash}X}
  
  \begin{tabularx}{\textwidth}{l|YY|YY|YY|YY}
    \hline
    \multicolumn{1}{l|}{\multirow{2}{*}{Variants}}
    &\multicolumn{2}{c}{\textbf{Movie}}&
    \multicolumn{2}{c|}{\textbf{Book}}&
    \multicolumn{2}{c}{\textbf{Food}}&\multicolumn{2}{c}{\textbf{Kitchen}}\\
    \cline{2-9} 
           &HR@5&NDCG@5&HR@5&NDCG@5&HR@5&NDCG@5&HR@5&NDCG@5\\
    \hline
    w/o Diffusion  & 0.0354 & 0.0210 & 0.0533 & 0.0288 & 0.2200 & 0.1368 & 0.2747 & 0.1911\\
    w/o Overlap Diffusion  & 0.0289 & 0.0178 & 0.0533 & 0.0351 & 0.0900 & 0.0584 & 0.0549 & 0.0353\\
    w/o Pseudo Diffusion & 0.0322 & 0.0150 & 0.0514 & 0.0316 & 0.2667 & 0.1778 & 0.2543 & 0.1741\\
    w/o Alignment Loss  & 0.0322 & 0.0194 & 0.0476 & 0.0334 & 0.2233 & 0.1430 & 0.2449 & 0.1583\\
    w/o Guesser Network & 0.0322 & 0.0209 & 0.0514 & 0.0316 & 0.2667 & 0.1798 & 0.2841 & 0.1925\\
    w/o MoE Fusion & 0.0129 & 0.0063 & 0.0343 & 0.0236 & 0.2665 & 0.1757 & 0.2575 &0.1730\\
    w/o Overlap Cyclic  & 0.0480 & \textbf{0.0309} & 0.0324 & 0.0205 & 0.2100 & 0.1305 & 0.2637 & 0.1833\\
    \hline
    \textbf{LGCD} & \textbf{0.0482} & 0.0286 & \textbf{0.0590} & \textbf{0.0373}& \textbf{0.2800} & \textbf{0.1805} & \textbf{0.3093}& \textbf{0.2140}\\
    \hline
  \end{tabularx}
\end{table*}

\subsection{Overall Comparison}
From the comparative experimental results presented in Table~\ref{tab:compare}, we draw the following conclusions:

(1) LGCD achieves the best performance across all metrics in all domains. This demonstrates that our proposed method is effectively applicable to CDR tasks.

(2) Single-domain methods outperform overlapping cross-domain methods (e.g., C2DSR, CTT) on many metrics. This is because the latter are designed exclusively for overlapping users, and incorporating single-domain sequences into their training disrupts the inter-domain relationships learned from overlapping sequences.

(3) Non-overlapping recommendation methods (e.g., DA-DAN, PLCR) achieve competitive performance across various datasets. These methods effectively leverage single-domain users to learn common inter-domain knowledge for recommending items in the target domain.

(4) Although LLMCDSR is designed to utilize both overlapping and non-overlapping sequences, it exhibits suboptimal performance. The primary reason is that LLMCDSR essentially employs LLMs and meta-learning to train a recommender for mixed cross-domain sequences. Consequently, when the interaction history consists of only a single domain, the model struggles to map it into the feature space of the other domain.

\subsection{Further Analysis}
\subsubsection{Ablation Studies}\label{sec:ablation}
To verify the contributions of individual components within LGCD, we compare it against the following ablation variants:
\begin{itemize}[leftmargin=*]
\item \textbf{w/o Diffusion:} Removes the diffusion model and directly employs conditional signals for prediction.
\item \textbf{w/o Overlap Diffusion:} Trains the diffusion model using only pseudo-overlapping interactions.
\item \textbf{w/o Pesudo Diffusion:} Trains the diffusion model using only real overlapping interactions.
\item \textbf{w/o Alignment Loss:} Removes the alignment constraint loss $\mathcal{L}_\text{align}$ applied to the pseudo-overlapping paths.
\item \textbf{w/o Guesser Network:} Uses a random vector as the initial representation during the reverse process instead of the prior generated by the guesser network.
\item \textbf{w/o MoE Fusion:} Replaces the MoE fusion with a simple linear fusion network to combine conditional signals and features restored by the diffusion model.
\item \textbf{w/o Overlap Cyclic:} Discards the cyclic batch construction for real overlapping interactions and adopts a standard data loading strategy.
\end{itemize}

From the comparative results in Table~\ref{tab:ablation}, we observe the following:

(1) Relying solely on source-domain features fails to adequately capture fine-grained user preferences in the target domain. Explicitly restoring target-domain features via the diffusion model significantly enhances cross-domain transfer capability. Furthermore, using exclusively real overlapping interactions or pseudo-overlapping interactions weakens the expressiveness of the diffusion model, indicating that these two types of overlapping signals are complementary and essential components during training.

(2)  Both supervision mechanisms we design the cyclic batch for overlapping interactions and the alignment loss, effectively mitigate the detriment caused by potential semantic noise and feature deviation inherent in pseudo-overlapping interaction data.

(3) The guesser network provides a reasonable target-domain prior for the diffusion process. This constrains the restoration trajectory within a subspace more consistent with target-domain semantics, facilitating the generation of more accurate and usable representations of target-domain preferences.

(4) The MoE-based fusion mechanism adaptively selects and weights different experts based on conditional signals. This approach simultaneously preserves stable preference patterns from the source domain and fully leverages the target-domain features generated by the diffusion model, thereby achieving superior cross-domain mapping and alignment of long-term user preferences.

\begin{figure}
\setlength{\abovecaptionskip}{0.1cm}
\setlength{\belowcaptionskip}{0.1cm} 
\centering
\includegraphics[width=8.5cm]{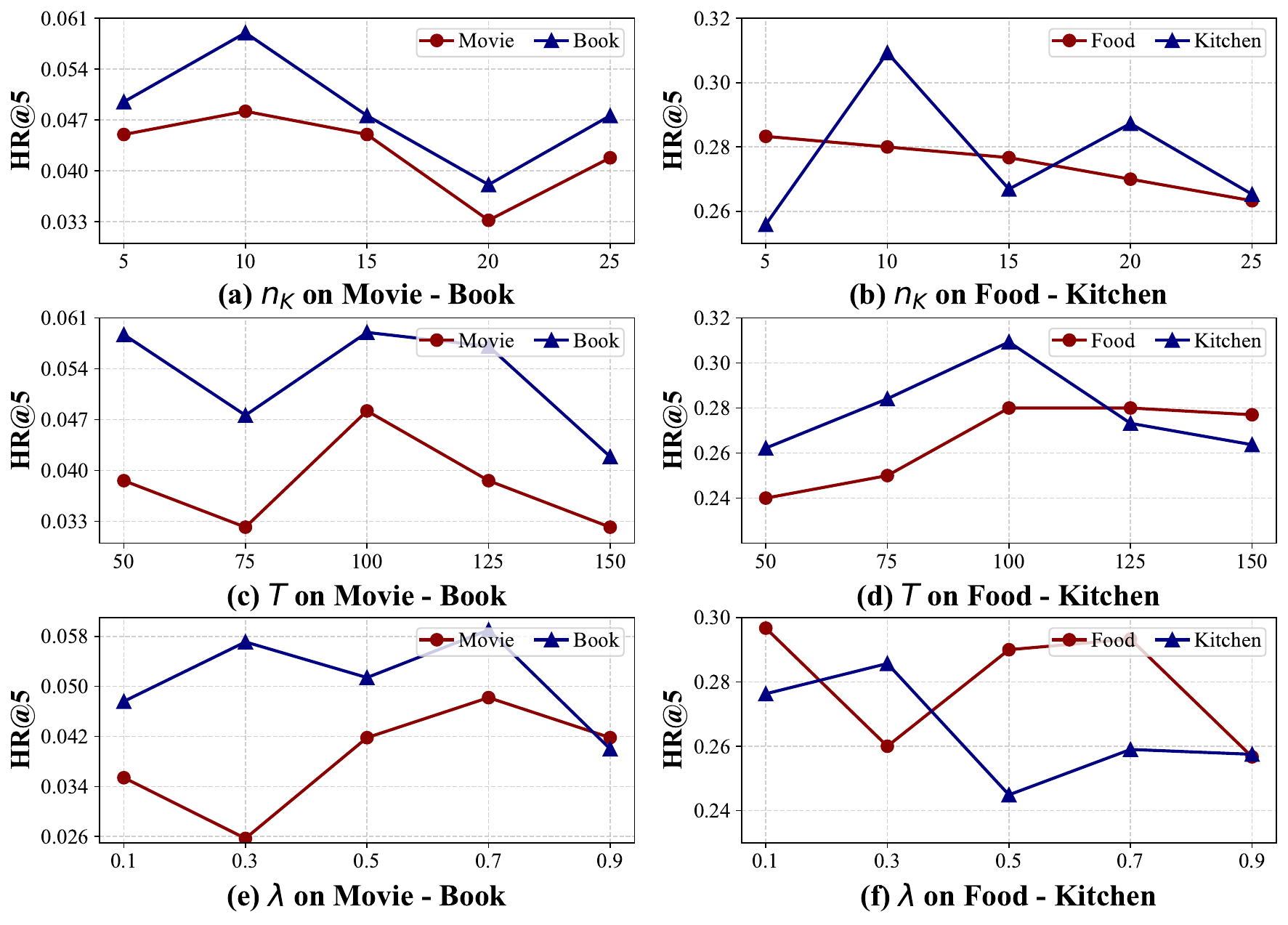}
\caption{Impact of the hyper-parameters $n_K$, $n_\mathrm{shared}$, $n_\mathrm{specific}$ on the Movie-Book and Food-Kitchen datasets.}
\label{fig:param}
\end{figure}
\subsubsection{Hyper-parameter Sensitivities}
To investigate the impact of hyperparameters, we analyze the model's sensitivity to three key parameters: the number of generated pseudo-items $n_K$, the number of diffusion generation steps $T$, and the start-step ratio $\lambda$. Fig.~\ref{fig:param} illustrates the performance variations in terms of HR@5 across different datasets. From these results, we observe the following: (1) The model achieves optimal performance on all datasets when $n_K$ is set to 10. As $n_K$ increases further, performance exhibits a downward trend. This indicates that an appropriate number of pseudo-items helps enrich cross-domain constraints, but too many pseudo-items can introduce more semantic noise, which interferes with the modeling of genuine user preferences. (2)  Performance initially improves and then degrades as the number of reverse generation steps increases, peaking at 100 steps. Too few steps result in an insufficient denoising process, failing to recover fine-grained target-domain features. Conversely, too many steps lead to over-smoothing and error accumulation, thereby diminishing the discriminative power of the final representation. (3) The model is relatively sensitive to $\lambda$. The optimal value is 0.7 for the Movie-Book dataset, 0.5 for the Food dataset, and 0.3 for the Kitchen dataset. This suggests that the optimal range of high-noise timesteps to retain varies across different cross-domain scenarios.
\begin{figure}
\setlength{\abovecaptionskip}{0.1cm}
\setlength{\belowcaptionskip}{0.1cm} 
\centering
\includegraphics[width=8.5cm]{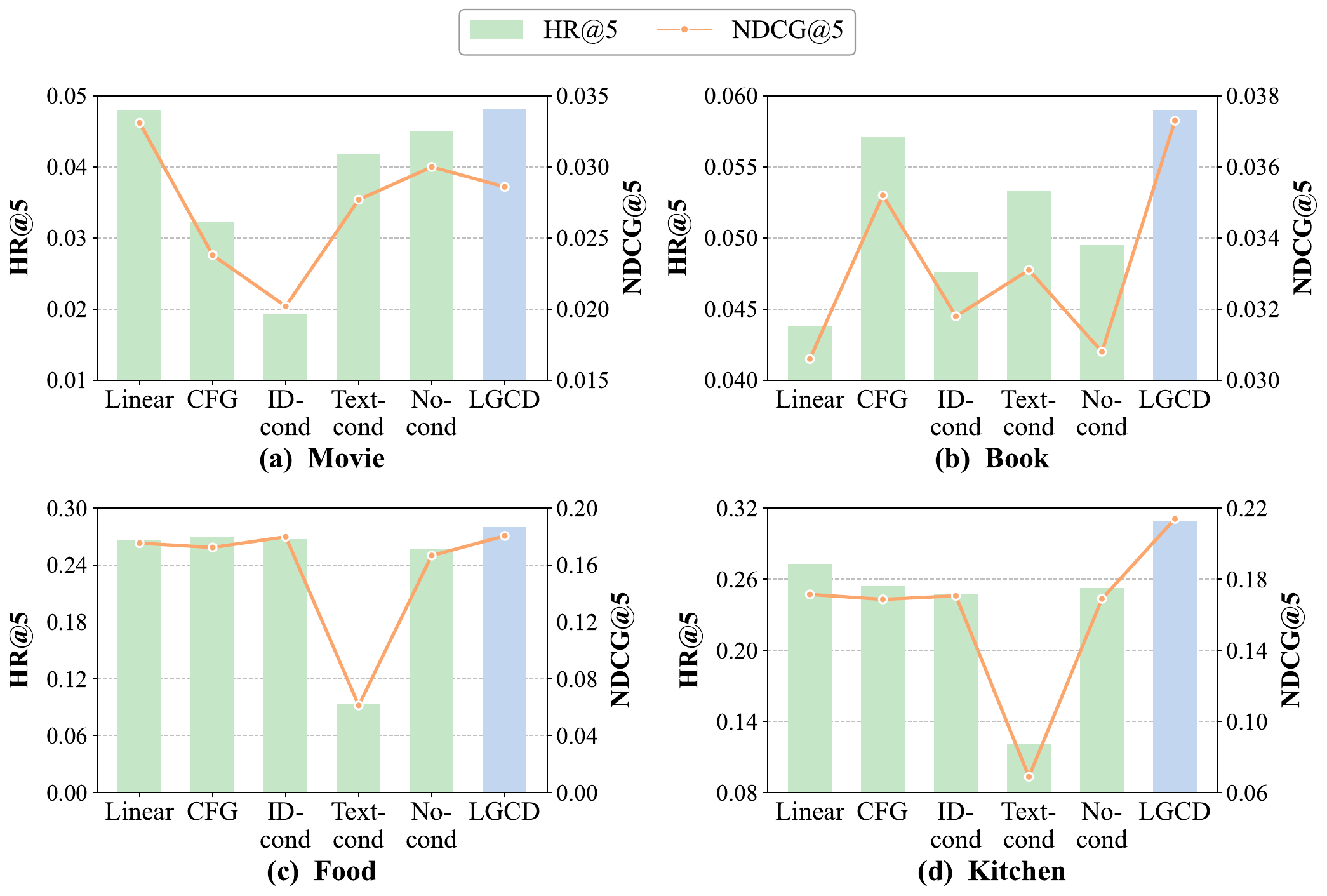}
\caption{Result of different conditional guidance methods on Movie-Book and Food-Kitchen datasets.}
\label{fig:cond}
\end{figure}

\subsubsection{Impact of Conditional Guidance Methods}\label{sec:exp_cond}
To identify the optimal conditional guidance strategy, we compare several conditional guidance mechanisms designed for the diffusion model. The results are presented in Fig.~\ref{fig:cond}. The specific variants include:
\begin{itemize}[leftmargin=*]
\item Linear: A simple linear fusion layer is used in the denoising network to fuse the conditional signal with the noisy features of the current step.
\item CFG: Classifier-Free Guidance~\cite{ho2022classifier} is adopted, combining conditional and unconditional predictions to derive the final denoising output.
\item ID-cond: Only the collaborative sequence representation is utilized as the conditional signal for denoising.
\item Text-cond: Only the semantic sequence representation is utilized as the conditional signal for denoising.
\item No-cond: No conditional signal is introduced, and the denoising network degenerates into a pure self-attention structure.
\end{itemize}
From the results, we observe that: (1) Compared to other baselines, LGCD achieves superior performance across most datasets and evaluation metrics. 
This demonstrates that cross-attention enables a more fine-grained alignment between conditional signals and noisy features, thereby providing more effective guidance for the diffusion process. 
(2) On the Movie–Book dataset, ID-cond exhibits the lowest overall performance, indicating a significant discrepancy in collaborative preferences between the two domains; relying solely on collaborative sequences is insufficient for reliable transfer. Conversely, on the Food–Kitchen dataset, Text-cond performs the worst, suggesting a substantial gap in the semantic spaces of these domains, where using text semantics alone makes it difficult for the diffusion model to accurately reconstruct target domain preferences. Collectively, these findings imply that single-type conditional signals fail to capture the diversity of cross-domain preferences, underscoring the necessity of jointly modeling collaborative and semantic information in LGCD.
\begin{figure}
\setlength{\abovecaptionskip}{0.1cm}
\setlength{\belowcaptionskip}{0.1cm} 
\centering
\includegraphics[width=8.5cm]{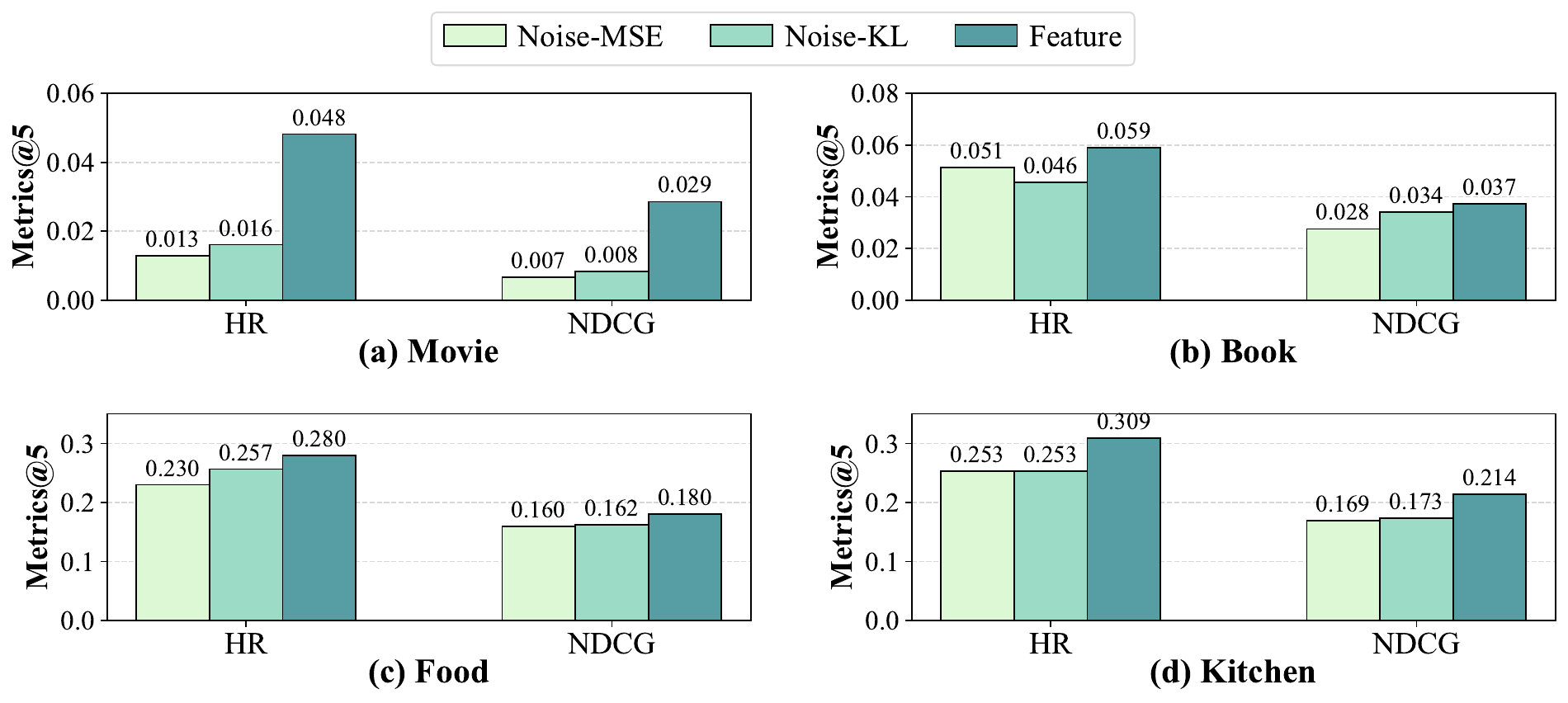}
\caption{The impact of different training targets for diffusion model on Movie-Book and Food-Kitchen}
\label{fig:pred}
\end{figure}

\begin{figure}
\setlength{\abovecaptionskip}{0.1cm}
\setlength{\belowcaptionskip}{0.1cm} 
\centering
\includegraphics[width=8.5cm]{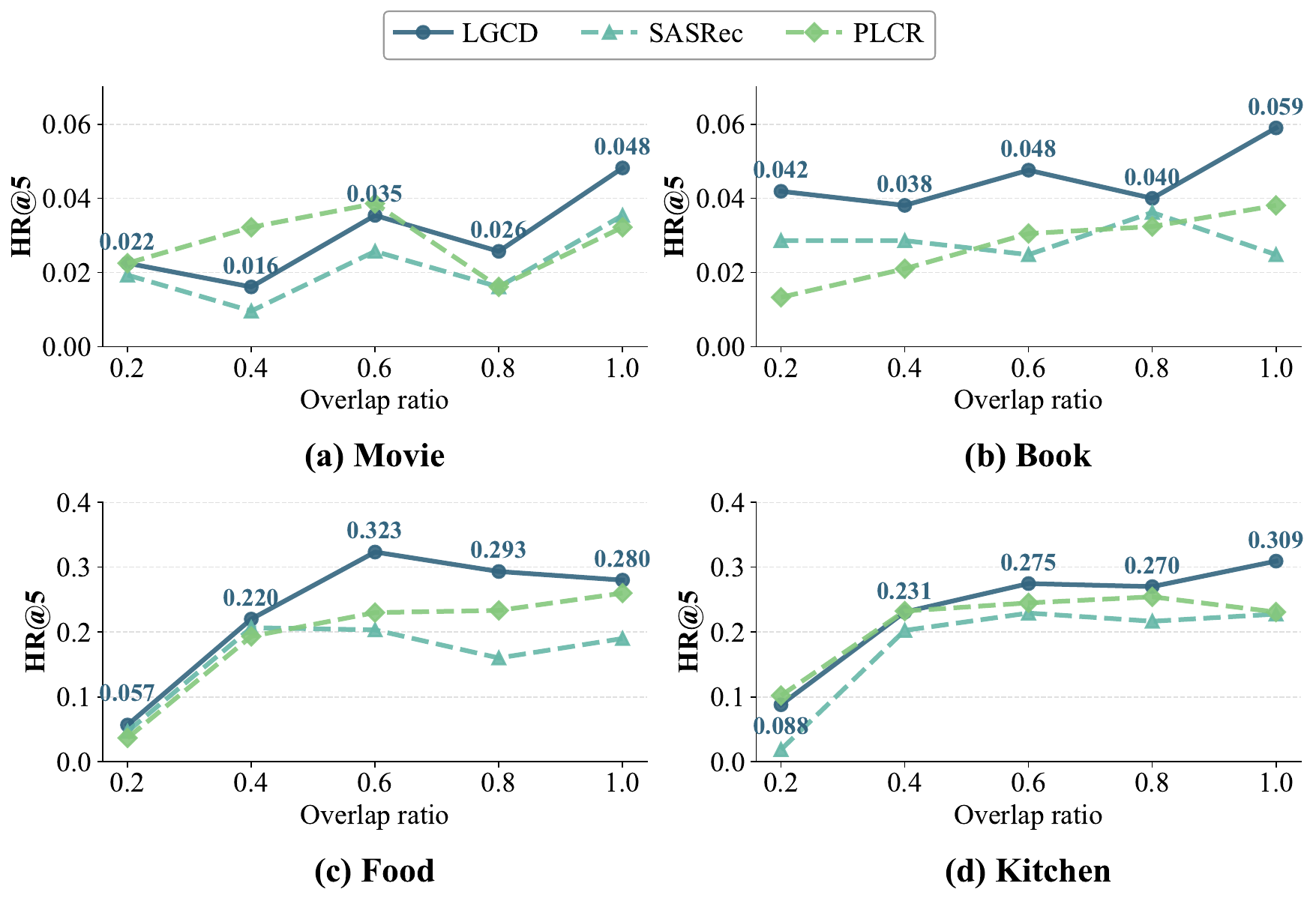}
\caption{The influence of overlap ratio on Movie-Book and Food-Kitchen datasets.}
\label{fig:overlap}
\end{figure}
\subsubsection{Study on Training Targets for Diffusion Models}\label{sec:diff_tgt}
To investigate the optimal training target for the diffusion model, we compare our approach against two variants that predict noise: Noise-MSE and Noise-KL, which employ MSE and KL divergence as the diffusion loss, respectively. The results in Fig.~\ref{fig:pred} demonstrate that directly predicting target features yields significantly better performance than predicting noise. This suggests that directly regressing semantic features is more suitable for our current task than fitting the noise distribution.

\subsubsection{Influence of Overlap Ratio}
To investigate the impact of real overlapping sequence scale on model performance, we vary the proportion of ground-truth overlapping data (overlap ratio) in the training set within the range of $[0.2, 1.0]$. As shown in Fig.~\ref{fig:overlap}, LGCD consistently outperforms the baseline methods across most overlap ratios.
When the overlap ratio is low (e.g., 0.2), the available overlapping interactions are insufficient, providing limited cross-domain supervisory signals, which results in suboptimal performance. As the overlap ratio increases, model performance gradually improves. Notably, performance tends to stabilize when the ratio reaches approximately 0.6. This indicates that at this threshold, the ground-truth overlapping sequences provide sufficient constraints for the model. Consequently, LGCD effectively leverages these reliable signals to better utilize pseudo-overlapping data, thereby enhancing the generative capability of the diffusion process.
\section{Conclusion}
In this work, we address the challenge of limited overlapping users in inter-domain recommendation by proposing LGCD, a novel framework that integrates Large Language Models with conditional diffusion models. We leverage LLMs to infer potential cross-domain preferences and map them to real items for valid pseudo-interaction construction. To facilitate precise knowledge transfer, we design a conditional diffusion architecture driven by cross-attention, which generates target user representations directly from source-domain conditions. Furthermore, we employ consistency constraints and a cyclic batching scheme to stabilize the alignment between real and generated distributions. Extensive experiments confirm that LGCD significantly outperforms state-of-the-art methods, providing an effective solution for cold-start users in cross-domain scenarios.


\begin{acks}

\end{acks}

\bibliographystyle{ACM-Reference-Format}
\balance
\bibliography{references}


\begin{thebibliography}{55}


\ifx \showCODEN    \undefined \def \showCODEN     #1{\unskip}     \fi
\ifx \showDOI      \undefined \def \showDOI       #1{#1}\fi
\ifx \showISBNx    \undefined \def \showISBNx     #1{\unskip}     \fi
\ifx \showISBNxiii \undefined \def \showISBNxiii  #1{\unskip}     \fi
\ifx \showISSN     \undefined \def \showISSN      #1{\unskip}     \fi
\ifx \showLCCN     \undefined \def \showLCCN      #1{\unskip}     \fi
\ifx \shownote     \undefined \def \shownote      #1{#1}          \fi
\ifx \showarticletitle \undefined \def \showarticletitle #1{#1}   \fi
\ifx \showURL      \undefined \def \showURL       {\relax}        \fi
\providecommand\bibfield[2]{#2}
\providecommand\bibinfo[2]{#2}
\providecommand\natexlab[1]{#1}
\providecommand\showeprint[2][]{arXiv:#2}

\bibitem[Cao et~al\mbox{.}(2022a)]%
        {cao2022contrastive}
\bibfield{author}{\bibinfo{person}{Jiangxia Cao}, \bibinfo{person}{Xin Cong},
  \bibinfo{person}{Jiawei Sheng}, \bibinfo{person}{Tingwen Liu}, {and}
  \bibinfo{person}{Bin Wang}.} \bibinfo{year}{2022}\natexlab{a}.
\newblock \showarticletitle{Contrastive cross-domain sequential
  recommendation}. In \bibinfo{booktitle}{\emph{Proceedings of the 31st ACM
  International Conference on Information \& Knowledge Management}}.
  \bibinfo{pages}{138--147}.
\newblock


\bibitem[Cao et~al\mbox{.}(2023)]%
        {cao2023towards}
\bibfield{author}{\bibinfo{person}{Jiangxia Cao}, \bibinfo{person}{Shaoshuai
  Li}, \bibinfo{person}{Bowen Yu}, \bibinfo{person}{Xiaobo Guo},
  \bibinfo{person}{Tingwen Liu}, {and} \bibinfo{person}{Bin Wang}.}
  \bibinfo{year}{2023}\natexlab{}.
\newblock \showarticletitle{Towards universal cross-domain recommendation}. In
  \bibinfo{booktitle}{\emph{Proceedings of the Sixteenth ACM International
  Conference on web search and data mining}}. \bibinfo{pages}{78--86}.
\newblock


\bibitem[Cao et~al\mbox{.}(2022b)]%
        {cao2022cross}
\bibfield{author}{\bibinfo{person}{Jiangxia Cao}, \bibinfo{person}{Jiawei
  Sheng}, \bibinfo{person}{Xin Cong}, \bibinfo{person}{Tingwen Liu}, {and}
  \bibinfo{person}{Bin Wang}.} \bibinfo{year}{2022}\natexlab{b}.
\newblock \showarticletitle{Cross-domain recommendation to cold-start users via
  variational information bottleneck}. In \bibinfo{booktitle}{\emph{2022 IEEE
  38th International Conference on data engineering (ICDE)}}. IEEE,
  \bibinfo{pages}{2209--2223}.
\newblock


\bibitem[Chen et~al\mbox{.}(2024)]%
        {chen2024survey}
\bibfield{author}{\bibinfo{person}{Shu Chen}, \bibinfo{person}{Zitao Xu},
  \bibinfo{person}{Weike Pan}, \bibinfo{person}{Qiang Yang}, {and}
  \bibinfo{person}{Zhong Ming}.} \bibinfo{year}{2024}\natexlab{}.
\newblock \showarticletitle{A survey on cross-domain sequential
  recommendation}.
\newblock \bibinfo{journal}{\emph{arXiv preprint arXiv:2401.04971}}
  (\bibinfo{year}{2024}).
\newblock


\bibitem[Chen et~al\mbox{.}(2025)]%
        {chen2025leave}
\bibfield{author}{\bibinfo{person}{Weixin Chen}, \bibinfo{person}{Yuhan Zhao},
  \bibinfo{person}{Li Chen}, {and} \bibinfo{person}{Weike Pan}.}
  \bibinfo{year}{2025}\natexlab{}.
\newblock \showarticletitle{Leave No One Behind: Fairness-Aware Cross-Domain
  Recommender Systems for Non-Overlapping Users}. In
  \bibinfo{booktitle}{\emph{Proceedings of the Nineteenth ACM Conference on
  Recommender Systems}}. \bibinfo{pages}{226--236}.
\newblock


\bibitem[Croitoru et~al\mbox{.}(2023)]%
        {croitoru2023diffusion}
\bibfield{author}{\bibinfo{person}{Florinel-Alin Croitoru},
  \bibinfo{person}{Vlad Hondru}, \bibinfo{person}{Radu~Tudor Ionescu}, {and}
  \bibinfo{person}{Mubarak Shah}.} \bibinfo{year}{2023}\natexlab{}.
\newblock \showarticletitle{Diffusion models in vision: A survey}.
\newblock \bibinfo{journal}{\emph{IEEE transactions on pattern analysis and
  machine intelligence}} \bibinfo{volume}{45}, \bibinfo{number}{9}
  (\bibinfo{year}{2023}), \bibinfo{pages}{10850--10869}.
\newblock


\bibitem[Du et~al\mbox{.}(2023)]%
        {du2023frequency}
\bibfield{author}{\bibinfo{person}{Xinyu Du}, \bibinfo{person}{Huanhuan Yuan},
  \bibinfo{person}{Pengpeng Zhao}, \bibinfo{person}{Jianfeng Qu},
  \bibinfo{person}{Fuzhen Zhuang}, \bibinfo{person}{Guanfeng Liu},
  \bibinfo{person}{Yanchi Liu}, {and} \bibinfo{person}{Victor~S Sheng}.}
  \bibinfo{year}{2023}\natexlab{}.
\newblock \showarticletitle{Frequency enhanced hybrid attention network for
  sequential recommendation}. In \bibinfo{booktitle}{\emph{Proceedings of the
  46th international ACM SIGIR conference on research and development in
  information retrieval}}. \bibinfo{pages}{78--88}.
\newblock


\bibitem[Guo et~al\mbox{.}(2023)]%
        {guo2023dan}
\bibfield{author}{\bibinfo{person}{Lei Guo}, \bibinfo{person}{Hao Liu},
  \bibinfo{person}{Lei Zhu}, \bibinfo{person}{Weili Guan}, {and}
  \bibinfo{person}{Zhiyong Cheng}.} \bibinfo{year}{2023}\natexlab{}.
\newblock \showarticletitle{DA-DAN: A dual adversarial domain adaption network
  for unsupervised non-overlapping cross-domain recommendation}.
\newblock \bibinfo{journal}{\emph{ACM Transactions on Information Systems}}
  \bibinfo{volume}{42}, \bibinfo{number}{2} (\bibinfo{year}{2023}),
  \bibinfo{pages}{1--27}.
\newblock


\bibitem[Guo et~al\mbox{.}(2025)]%
        {guo2025automated}
\bibfield{author}{\bibinfo{person}{Lei Guo}, \bibinfo{person}{Chunxiao Wang},
  \bibinfo{person}{Xinhua Wang}, \bibinfo{person}{Lei Zhu}, {and}
  \bibinfo{person}{Hongzhi Yin}.} \bibinfo{year}{2025}\natexlab{}.
\newblock \showarticletitle{Automated prompting for non-overlapping
  cross-domain sequential recommendation}.
\newblock \bibinfo{journal}{\emph{IEEE Transactions on Knowledge and Data
  Engineering}} (\bibinfo{year}{2025}).
\newblock


\bibitem[Hidasi et~al\mbox{.}(2015)]%
        {hidasi2015session}
\bibfield{author}{\bibinfo{person}{Bal{\'a}zs Hidasi},
  \bibinfo{person}{Alexandros Karatzoglou}, \bibinfo{person}{Linas Baltrunas},
  {and} \bibinfo{person}{Domonkos Tikk}.} \bibinfo{year}{2015}\natexlab{}.
\newblock \showarticletitle{Session-based recommendations with recurrent neural
  networks}.
\newblock \bibinfo{journal}{\emph{arXiv preprint arXiv:1511.06939}}
  (\bibinfo{year}{2015}).
\newblock


\bibitem[Ho and Salimans(2022)]%
        {ho2022classifier}
\bibfield{author}{\bibinfo{person}{Jonathan Ho} {and} \bibinfo{person}{Tim
  Salimans}.} \bibinfo{year}{2022}\natexlab{}.
\newblock \showarticletitle{Classifier-free diffusion guidance}.
\newblock \bibinfo{journal}{\emph{arXiv preprint arXiv:2207.12598}}
  (\bibinfo{year}{2022}).
\newblock


\bibitem[Hou et~al\mbox{.}(2022)]%
        {hou2022towards}
\bibfield{author}{\bibinfo{person}{Yupeng Hou}, \bibinfo{person}{Shanlei Mu},
  \bibinfo{person}{Wayne~Xin Zhao}, \bibinfo{person}{Yaliang Li},
  \bibinfo{person}{Bolin Ding}, {and} \bibinfo{person}{Ji-Rong Wen}.}
  \bibinfo{year}{2022}\natexlab{}.
\newblock \showarticletitle{Towards universal sequence representation learning
  for recommender systems}. In \bibinfo{booktitle}{\emph{Proceedings of the
  28th ACM SIGKDD conference on knowledge discovery and data mining}}.
  \bibinfo{pages}{585--593}.
\newblock


\bibitem[Huang et~al\mbox{.}(2025)]%
        {huang2025survey}
\bibfield{author}{\bibinfo{person}{Lei Huang}, \bibinfo{person}{Weijiang Yu},
  \bibinfo{person}{Weitao Ma}, \bibinfo{person}{Weihong Zhong},
  \bibinfo{person}{Zhangyin Feng}, \bibinfo{person}{Haotian Wang},
  \bibinfo{person}{Qianglong Chen}, \bibinfo{person}{Weihua Peng},
  \bibinfo{person}{Xiaocheng Feng}, \bibinfo{person}{Bing Qin},
  {et~al\mbox{.}}} \bibinfo{year}{2025}\natexlab{}.
\newblock \showarticletitle{A survey on hallucination in large language models:
  Principles, taxonomy, challenges, and open questions}.
\newblock \bibinfo{journal}{\emph{ACM Transactions on Information Systems}}
  \bibinfo{volume}{43}, \bibinfo{number}{2} (\bibinfo{year}{2025}),
  \bibinfo{pages}{1--55}.
\newblock


\bibitem[Jin et~al\mbox{.}(2025)]%
        {jin2025diffusion}
\bibfield{author}{\bibinfo{person}{Ke Jin}, \bibinfo{person}{Weihao Yu},
  \bibinfo{person}{Yingchao Long}, \bibinfo{person}{Nanhui Lai}, {and}
  \bibinfo{person}{Jin Huang}.} \bibinfo{year}{2025}\natexlab{}.
\newblock \showarticletitle{A diffusion multi-interest framework for
  cross-domain recommendation}.
\newblock \bibinfo{journal}{\emph{Expert Systems with Applications}}
  (\bibinfo{year}{2025}), \bibinfo{pages}{128738}.
\newblock


\bibitem[Kang et~al\mbox{.}(2019)]%
        {kang2019semi}
\bibfield{author}{\bibinfo{person}{SeongKu Kang}, \bibinfo{person}{Junyoung
  Hwang}, \bibinfo{person}{Dongha Lee}, {and} \bibinfo{person}{Hwanjo Yu}.}
  \bibinfo{year}{2019}\natexlab{}.
\newblock \showarticletitle{Semi-supervised learning for cross-domain
  recommendation to cold-start users}. In \bibinfo{booktitle}{\emph{Proceedings
  of the 28th ACM international conference on information and knowledge
  management}}. \bibinfo{pages}{1563--1572}.
\newblock


\bibitem[Kang and McAuley(2018)]%
        {kang2018self}
\bibfield{author}{\bibinfo{person}{Wang-Cheng Kang} {and}
  \bibinfo{person}{Julian McAuley}.} \bibinfo{year}{2018}\natexlab{}.
\newblock \showarticletitle{Self-attentive sequential recommendation}. In
  \bibinfo{booktitle}{\emph{2018 IEEE international conference on data mining
  (ICDM)}}. IEEE, \bibinfo{pages}{197--206}.
\newblock


\bibitem[Kim et~al\mbox{.}(2025)]%
        {kim2025lost}
\bibfield{author}{\bibinfo{person}{Sein Kim}, \bibinfo{person}{Hongseok Kang},
  \bibinfo{person}{Kibum Kim}, \bibinfo{person}{Jiwan Kim},
  \bibinfo{person}{Donghyun Kim}, \bibinfo{person}{Minchul Yang},
  \bibinfo{person}{Kwangjin Oh}, \bibinfo{person}{Julian McAuley}, {and}
  \bibinfo{person}{Chanyoung Park}.} \bibinfo{year}{2025}\natexlab{}.
\newblock \showarticletitle{Lost in Sequence: Do Large Language Models
  Understand Sequential Recommendation?}. In
  \bibinfo{booktitle}{\emph{Proceedings of the 31st ACM SIGKDD Conference on
  Knowledge Discovery and Data Mining V. 2}}. \bibinfo{pages}{1160--1171}.
\newblock


\bibitem[Kingma(2014)]%
        {kingma2014adam}
\bibfield{author}{\bibinfo{person}{Diederik~P Kingma}.}
  \bibinfo{year}{2014}\natexlab{}.
\newblock \showarticletitle{Adam: A method for stochastic optimization}.
\newblock \bibinfo{journal}{\emph{arXiv preprint arXiv:1412.6980}}
  (\bibinfo{year}{2014}).
\newblock


\bibitem[Li et~al\mbox{.}(2025b)]%
        {li2025diffusion}
\bibfield{author}{\bibinfo{person}{Fengxin Li}, \bibinfo{person}{Hongyan Liu},
  {and} \bibinfo{person}{Jun He}.} \bibinfo{year}{2025}\natexlab{b}.
\newblock \showarticletitle{Diffusion Alignment for Cross Domain
  Recommendation}. In \bibinfo{booktitle}{\emph{Proceedings of the 2025
  International Conference on Multimedia Retrieval}}.
  \bibinfo{pages}{715--723}.
\newblock


\bibitem[Li et~al\mbox{.}(2025a)]%
        {li2025cd}
\bibfield{author}{\bibinfo{person}{Hanyu Li}, \bibinfo{person}{Jiayu Li},
  \bibinfo{person}{Weizhi Ma}, \bibinfo{person}{Peijie Sun},
  \bibinfo{person}{Haiyang Wu}, \bibinfo{person}{Jingwen Wang},
  \bibinfo{person}{Yuekui Yang}, \bibinfo{person}{Min Zhang}, {and}
  \bibinfo{person}{Shaoping Ma}.} \bibinfo{year}{2025}\natexlab{a}.
\newblock \showarticletitle{CD-CDR: Conditional Diffusion-based Item Generation
  for Cross-Domain Recommendation}. In \bibinfo{booktitle}{\emph{Proceedings of
  the 48th International ACM SIGIR Conference on Research and Development in
  Information Retrieval}}. \bibinfo{pages}{1789--1798}.
\newblock


\bibitem[Li et~al\mbox{.}(2024b)]%
        {li2024aiming}
\bibfield{author}{\bibinfo{person}{Hanyu Li}, \bibinfo{person}{Weizhi Ma},
  \bibinfo{person}{Peijie Sun}, \bibinfo{person}{Jiayu Li},
  \bibinfo{person}{Cunxiang Yin}, \bibinfo{person}{Yancheng He},
  \bibinfo{person}{Guoqiang Xu}, \bibinfo{person}{Min Zhang}, {and}
  \bibinfo{person}{Shaoping Ma}.} \bibinfo{year}{2024}\natexlab{b}.
\newblock \showarticletitle{Aiming at the target: Filter collaborative
  information for cross-domain recommendation}. In
  \bibinfo{booktitle}{\emph{Proceedings of the 47th International ACM SIGIR
  Conference on Research and Development in Information Retrieval}}.
  \bibinfo{pages}{2081--2090}.
\newblock


\bibitem[Li et~al\mbox{.}(2025d)]%
        {li2025disco}
\bibfield{author}{\bibinfo{person}{Hourun Li}, \bibinfo{person}{Yifan Wang},
  \bibinfo{person}{Zhiping Xiao}, \bibinfo{person}{Jia Yang},
  \bibinfo{person}{Changling Zhou}, \bibinfo{person}{Ming Zhang}, {and}
  \bibinfo{person}{Wei Ju}.} \bibinfo{year}{2025}\natexlab{d}.
\newblock \showarticletitle{DisCo: graph-based disentangled contrastive
  learning for cold-start cross-domain recommendation}. In
  \bibinfo{booktitle}{\emph{Proceedings of the AAAI Conference on Artificial
  Intelligence}}, Vol.~\bibinfo{volume}{39}. \bibinfo{pages}{12049--12057}.
\newblock


\bibitem[Li et~al\mbox{.}(2024d)]%
        {li2024large}
\bibfield{author}{\bibinfo{person}{Lei Li}, \bibinfo{person}{Yongfeng Zhang},
  \bibinfo{person}{Dugang Liu}, {and} \bibinfo{person}{Li Chen}.}
  \bibinfo{year}{2024}\natexlab{d}.
\newblock \showarticletitle{Large language models for generative
  recommendation: A survey and visionary discussions}. In
  \bibinfo{booktitle}{\emph{Proceedings of the 2024 Joint International
  Conference on Computational Linguistics, Language Resources and Evaluation
  (LREC-COLING 2024)}}. \bibinfo{pages}{10146--10159}.
\newblock


\bibitem[Li and Tuzhilin(2020)]%
        {li2020ddtcdr}
\bibfield{author}{\bibinfo{person}{Pan Li} {and} \bibinfo{person}{Alexander
  Tuzhilin}.} \bibinfo{year}{2020}\natexlab{}.
\newblock \showarticletitle{Ddtcdr: Deep dual transfer cross domain
  recommendation}. In \bibinfo{booktitle}{\emph{Proceedings of the 13th
  international conference on web search and data mining}}.
  \bibinfo{pages}{331--339}.
\newblock


\bibitem[Li and He(2025)]%
        {li2025back}
\bibfield{author}{\bibinfo{person}{Tianhong Li} {and} \bibinfo{person}{Kaiming
  He}.} \bibinfo{year}{2025}\natexlab{}.
\newblock \showarticletitle{Back to basics: Let denoising generative models
  denoise}.
\newblock \bibinfo{journal}{\emph{arXiv preprint arXiv:2511.13720}}
  (\bibinfo{year}{2025}).
\newblock


\bibitem[Li et~al\mbox{.}(2024c)]%
        {li2024cdrnp}
\bibfield{author}{\bibinfo{person}{Xiaodong Li}, \bibinfo{person}{Jiawei
  Sheng}, \bibinfo{person}{Jiangxia Cao}, \bibinfo{person}{Wenyuan Zhang},
  \bibinfo{person}{Quangang Li}, {and} \bibinfo{person}{Tingwen Liu}.}
  \bibinfo{year}{2024}\natexlab{c}.
\newblock \showarticletitle{Cdrnp: Cross-domain recommendation to cold-start
  users via neural process}. In \bibinfo{booktitle}{\emph{Proceedings of the
  17th ACM international conference on web search and data mining}}.
  \bibinfo{pages}{378--386}.
\newblock


\bibitem[Li et~al\mbox{.}(2025c)]%
        {li2025exploring}
\bibfield{author}{\bibinfo{person}{Xiaodong Li}, \bibinfo{person}{Hengzhu
  Tang}, \bibinfo{person}{Jiawei Sheng}, \bibinfo{person}{Xinghua Zhang},
  \bibinfo{person}{Li Gao}, \bibinfo{person}{Suqi Cheng},
  \bibinfo{person}{Dawei Yin}, {and} \bibinfo{person}{Tingwen Liu}.}
  \bibinfo{year}{2025}\natexlab{c}.
\newblock \showarticletitle{Exploring Preference-Guided Diffusion Model for
  Cross-Domain Recommendation}.
\newblock \bibinfo{journal}{\emph{arXiv preprint arXiv:2501.11671}}
  (\bibinfo{year}{2025}).
\newblock


\bibitem[Li et~al\mbox{.}(2024a)]%
        {li2024mutual}
\bibfield{author}{\bibinfo{person}{Zhi Li}, \bibinfo{person}{Daichi Amagata},
  \bibinfo{person}{Yihong Zhang}, \bibinfo{person}{Takahiro Hara},
  \bibinfo{person}{Shuichiro Haruta}, \bibinfo{person}{Kei Yonekawa}, {and}
  \bibinfo{person}{Mori Kurokawa}.} \bibinfo{year}{2024}\natexlab{a}.
\newblock \showarticletitle{Mutual information-based preference disentangling
  and transferring for non-overlapped multi-target cross-domain
  recommendations}. In \bibinfo{booktitle}{\emph{Proceedings of the 47th
  International ACM SIGIR Conference on Research and Development in Information
  Retrieval}}. \bibinfo{pages}{2124--2133}.
\newblock


\bibitem[Li et~al\mbox{.}(2023)]%
        {li2023diffurec}
\bibfield{author}{\bibinfo{person}{Zihao Li}, \bibinfo{person}{Aixin Sun},
  {and} \bibinfo{person}{Chenliang Li}.} \bibinfo{year}{2023}\natexlab{}.
\newblock \showarticletitle{Diffurec: A diffusion model for sequential
  recommendation}.
\newblock \bibinfo{journal}{\emph{ACM Transactions on Information Systems}}
  \bibinfo{volume}{42}, \bibinfo{number}{3} (\bibinfo{year}{2023}),
  \bibinfo{pages}{1--28}.
\newblock


\bibitem[Lin et~al\mbox{.}(2024)]%
        {lin2024mixed}
\bibfield{author}{\bibinfo{person}{Guanyu Lin}, \bibinfo{person}{Chen Gao},
  \bibinfo{person}{Yu Zheng}, \bibinfo{person}{Jianxin Chang},
  \bibinfo{person}{Yanan Niu}, \bibinfo{person}{Yang Song},
  \bibinfo{person}{Kun Gai}, \bibinfo{person}{Zhiheng Li},
  \bibinfo{person}{Depeng Jin}, \bibinfo{person}{Yong Li}, {et~al\mbox{.}}}
  \bibinfo{year}{2024}\natexlab{}.
\newblock \showarticletitle{Mixed attention network for cross-domain sequential
  recommendation}. In \bibinfo{booktitle}{\emph{Proceedings of the 17th ACM
  international conference on web search and data mining}}.
  \bibinfo{pages}{405--413}.
\newblock


\bibitem[Liu et~al\mbox{.}(2024)]%
        {liu2024mcrpl}
\bibfield{author}{\bibinfo{person}{Hao Liu}, \bibinfo{person}{Lei Guo},
  \bibinfo{person}{Lei Zhu}, \bibinfo{person}{Yongqiang Jiang},
  \bibinfo{person}{Min Gao}, {and} \bibinfo{person}{Hongzhi Yin}.}
  \bibinfo{year}{2024}\natexlab{}.
\newblock \showarticletitle{MCRPL: A Pretrain, prompt, and fine-tune paradigm
  for non-overlapping many-to-one cross-domain recommendation}.
\newblock \bibinfo{journal}{\emph{ACM Transactions on Information Systems}}
  \bibinfo{volume}{42}, \bibinfo{number}{4} (\bibinfo{year}{2024}),
  \bibinfo{pages}{1--24}.
\newblock


\bibitem[Liu et~al\mbox{.}(2025)]%
        {liu2025llm}
\bibfield{author}{\bibinfo{person}{Kuan Liu}, \bibinfo{person}{Ke Wang},
  \bibinfo{person}{Ji Zhang}, {and} \bibinfo{person}{Gang Zhou}.}
  \bibinfo{year}{2025}\natexlab{}.
\newblock \showarticletitle{LLM-Grounded Diffusion for Cross-Domain
  Recommendation}. In \bibinfo{booktitle}{\emph{Proceedings of the 33rd ACM
  International Conference on Multimedia}}. \bibinfo{pages}{6103--6112}.
\newblock


\bibitem[Liu et~al\mbox{.}(2023)]%
        {liu2023diffusion}
\bibfield{author}{\bibinfo{person}{Qidong Liu}, \bibinfo{person}{Fan Yan},
  \bibinfo{person}{Xiangyu Zhao}, \bibinfo{person}{Zhaocheng Du},
  \bibinfo{person}{Huifeng Guo}, \bibinfo{person}{Ruiming Tang}, {and}
  \bibinfo{person}{Feng Tian}.} \bibinfo{year}{2023}\natexlab{}.
\newblock \showarticletitle{Diffusion augmentation for sequential
  recommendation}. In \bibinfo{booktitle}{\emph{Proceedings of the 32nd ACM
  International conference on information and knowledge management}}.
  \bibinfo{pages}{1576--1586}.
\newblock


\bibitem[Ma et~al\mbox{.}(2024a)]%
        {ma2024plug}
\bibfield{author}{\bibinfo{person}{Haokai Ma}, \bibinfo{person}{Ruobing Xie},
  \bibinfo{person}{Lei Meng}, \bibinfo{person}{Xin Chen}, \bibinfo{person}{Xu
  Zhang}, \bibinfo{person}{Leyu Lin}, {and} \bibinfo{person}{Zhanhui Kang}.}
  \bibinfo{year}{2024}\natexlab{a}.
\newblock \showarticletitle{Plug-in diffusion model for sequential
  recommendation}. In \bibinfo{booktitle}{\emph{Proceedings of the AAAI
  conference on artificial intelligence}}, Vol.~\bibinfo{volume}{38}.
  \bibinfo{pages}{8886--8894}.
\newblock


\bibitem[Ma et~al\mbox{.}(2024b)]%
        {ma2024triple}
\bibfield{author}{\bibinfo{person}{Haokai Ma}, \bibinfo{person}{Ruobing Xie},
  \bibinfo{person}{Lei Meng}, \bibinfo{person}{Xin Chen}, \bibinfo{person}{Xu
  Zhang}, \bibinfo{person}{Leyu Lin}, {and} \bibinfo{person}{Jie Zhou}.}
  \bibinfo{year}{2024}\natexlab{b}.
\newblock \showarticletitle{Triple sequence learning for cross-domain
  recommendation}.
\newblock \bibinfo{journal}{\emph{ACM Transactions on Information Systems}}
  \bibinfo{volume}{42}, \bibinfo{number}{4} (\bibinfo{year}{2024}),
  \bibinfo{pages}{1--29}.
\newblock


\bibitem[Man et~al\mbox{.}(2017)]%
        {man2017cross}
\bibfield{author}{\bibinfo{person}{Tong Man}, \bibinfo{person}{Huawei Shen},
  \bibinfo{person}{Xiaolong Jin}, {and} \bibinfo{person}{Xueqi Cheng}.}
  \bibinfo{year}{2017}\natexlab{}.
\newblock \showarticletitle{Cross-domain recommendation: An embedding and
  mapping approach.}. In \bibinfo{booktitle}{\emph{Ijcai}},
  Vol.~\bibinfo{volume}{17}. \bibinfo{pages}{2464--2470}.
\newblock


\bibitem[Rombach et~al\mbox{.}(2022)]%
        {rombach2022high}
\bibfield{author}{\bibinfo{person}{Robin Rombach}, \bibinfo{person}{Andreas
  Blattmann}, \bibinfo{person}{Dominik Lorenz}, \bibinfo{person}{Patrick
  Esser}, {and} \bibinfo{person}{Bj{\"o}rn Ommer}.}
  \bibinfo{year}{2022}\natexlab{}.
\newblock \showarticletitle{High-resolution image synthesis with latent
  diffusion models}. In \bibinfo{booktitle}{\emph{Proceedings of the IEEE/CVF
  conference on computer vision and pattern recognition}}.
  \bibinfo{pages}{10684--10695}.
\newblock


\bibitem[Tonmoy et~al\mbox{.}(2024)]%
        {tonmoy2024comprehensive}
\bibfield{author}{\bibinfo{person}{SMTI Tonmoy}, \bibinfo{person}{SM Zaman},
  \bibinfo{person}{Vinija Jain}, \bibinfo{person}{Anku Rani},
  \bibinfo{person}{Vipula Rawte}, \bibinfo{person}{Aman Chadha}, {and}
  \bibinfo{person}{Amitava Das}.} \bibinfo{year}{2024}\natexlab{}.
\newblock \showarticletitle{A comprehensive survey of hallucination mitigation
  techniques in large language models}.
\newblock \bibinfo{journal}{\emph{arXiv preprint arXiv:2401.01313}}
  \bibinfo{volume}{6} (\bibinfo{year}{2024}).
\newblock


\bibitem[Ulhaq and Akhtar(2022)]%
        {ulhaq2022efficient}
\bibfield{author}{\bibinfo{person}{Anwaar Ulhaq} {and} \bibinfo{person}{Naveed
  Akhtar}.} \bibinfo{year}{2022}\natexlab{}.
\newblock \showarticletitle{Efficient diffusion models for vision: A survey}.
\newblock \bibinfo{journal}{\emph{arXiv preprint arXiv:2210.09292}}
  (\bibinfo{year}{2022}).
\newblock


\bibitem[Vaswani et~al\mbox{.}(2017)]%
        {vaswani2017attention}
\bibfield{author}{\bibinfo{person}{Ashish Vaswani}, \bibinfo{person}{Noam
  Shazeer}, \bibinfo{person}{Niki Parmar}, \bibinfo{person}{Jakob Uszkoreit},
  \bibinfo{person}{Llion Jones}, \bibinfo{person}{Aidan~N Gomez},
  \bibinfo{person}{{\L}ukasz Kaiser}, {and} \bibinfo{person}{Illia
  Polosukhin}.} \bibinfo{year}{2017}\natexlab{}.
\newblock \showarticletitle{Attention is all you need}.
\newblock \bibinfo{journal}{\emph{Advances in neural information processing
  systems}}  \bibinfo{volume}{30} (\bibinfo{year}{2017}).
\newblock


\bibitem[Wang et~al\mbox{.}(2024a)]%
        {wang2024conditional}
\bibfield{author}{\bibinfo{person}{Yu Wang}, \bibinfo{person}{Zhiwei Liu},
  \bibinfo{person}{Liangwei Yang}, {and} \bibinfo{person}{Philip~S Yu}.}
  \bibinfo{year}{2024}\natexlab{a}.
\newblock \showarticletitle{Conditional denoising diffusion for sequential
  recommendation}. In \bibinfo{booktitle}{\emph{Pacific-Asia conference on
  knowledge discovery and data mining}}. Springer, \bibinfo{pages}{156--169}.
\newblock


\bibitem[Wang et~al\mbox{.}(2024b)]%
        {wang2024making}
\bibfield{author}{\bibinfo{person}{Zihan Wang}, \bibinfo{person}{Yonghui Yang},
  \bibinfo{person}{Le Wu}, \bibinfo{person}{Richang Hong}, {and}
  \bibinfo{person}{Meng Wang}.} \bibinfo{year}{2024}\natexlab{b}.
\newblock \showarticletitle{Making Non-Overlapping Matters: An Unsupervised
  Alignment Enhanced Cross-Domain Cold-Start Recommendation}.
\newblock \bibinfo{journal}{\emph{IEEE Transactions on Knowledge and Data
  Engineering}} (\bibinfo{year}{2024}).
\newblock


\bibitem[Wu et~al\mbox{.}(2024)]%
        {wu2024survey}
\bibfield{author}{\bibinfo{person}{Likang Wu}, \bibinfo{person}{Zhi Zheng},
  \bibinfo{person}{Zhaopeng Qiu}, \bibinfo{person}{Hao Wang},
  \bibinfo{person}{Hongchao Gu}, \bibinfo{person}{Tingjia Shen},
  \bibinfo{person}{Chuan Qin}, \bibinfo{person}{Chen Zhu},
  \bibinfo{person}{Hengshu Zhu}, \bibinfo{person}{Qi Liu}, {et~al\mbox{.}}}
  \bibinfo{year}{2024}\natexlab{}.
\newblock \showarticletitle{A survey on large language models for
  recommendation}.
\newblock \bibinfo{journal}{\emph{World Wide Web}} \bibinfo{volume}{27},
  \bibinfo{number}{5} (\bibinfo{year}{2024}), \bibinfo{pages}{60}.
\newblock


\bibitem[Xin et~al\mbox{.}(2025)]%
        {xin2025llmcdsr}
\bibfield{author}{\bibinfo{person}{Haoran Xin}, \bibinfo{person}{Ying Sun},
  \bibinfo{person}{Chao Wang}, {and} \bibinfo{person}{Hui Xiong}.}
  \bibinfo{year}{2025}\natexlab{}.
\newblock \showarticletitle{Llmcdsr: Enhancing cross-domain sequential
  recommendation with large language models}.
\newblock \bibinfo{journal}{\emph{ACM Transactions on Information Systems}}
  (\bibinfo{year}{2025}).
\newblock


\bibitem[Xu et~al\mbox{.}(2025)]%
        {xu2025multi}
\bibfield{author}{\bibinfo{person}{Zitao Xu}, \bibinfo{person}{Shu Chen},
  \bibinfo{person}{Weike Pan}, {and} \bibinfo{person}{Zhong Ming}.}
  \bibinfo{year}{2025}\natexlab{}.
\newblock \showarticletitle{A multi-view graph contrastive learning framework
  for cross-domain sequential recommendation}.
\newblock \bibinfo{journal}{\emph{ACM Transactions on Recommender Systems}}
  \bibinfo{volume}{3}, \bibinfo{number}{4} (\bibinfo{year}{2025}),
  \bibinfo{pages}{1--28}.
\newblock


\bibitem[Xuan(2024)]%
        {xuan2024diffusion}
\bibfield{author}{\bibinfo{person}{Yuner Xuan}.}
  \bibinfo{year}{2024}\natexlab{}.
\newblock \showarticletitle{Diffusion cross-domain recommendation}.
\newblock \bibinfo{journal}{\emph{arXiv preprint arXiv:2402.02182}}
  (\bibinfo{year}{2024}).
\newblock


\bibitem[Yang et~al\mbox{.}(2024b)]%
        {yang2024cross}
\bibfield{author}{\bibinfo{person}{Ling Yang}, \bibinfo{person}{Zhilong Zhang},
  \bibinfo{person}{Zhaochen Yu}, \bibinfo{person}{Jingwei Liu},
  \bibinfo{person}{Minkai Xu}, \bibinfo{person}{Stefano Ermon}, {and}
  \bibinfo{person}{Bin Cui}.} \bibinfo{year}{2024}\natexlab{b}.
\newblock \showarticletitle{Cross-modal contextualized diffusion models for
  text-guided visual generation and editing}. In \bibinfo{booktitle}{\emph{The
  Twelfth International Conference on Learning Representations}}.
\newblock


\bibitem[Yang et~al\mbox{.}(2024a)]%
        {yang2024not}
\bibfield{author}{\bibinfo{person}{Wenhao Yang}, \bibinfo{person}{Yingchun
  Jian}, \bibinfo{person}{Yibo Wang}, \bibinfo{person}{Shiyin Lu},
  \bibinfo{person}{Lei Shen}, \bibinfo{person}{Bing Wang},
  \bibinfo{person}{Haihong Tang}, {and} \bibinfo{person}{Lijun Zhang}.}
  \bibinfo{year}{2024}\natexlab{a}.
\newblock \showarticletitle{Not all embeddings are created equal: towards
  robust cross-domain recommendation via contrastive learning}. In
  \bibinfo{booktitle}{\emph{Proceedings of the ACM Web Conference 2024}}.
  \bibinfo{pages}{3195--3206}.
\newblock


\bibitem[Yang et~al\mbox{.}(2023)]%
        {yang2023generate}
\bibfield{author}{\bibinfo{person}{Zhengyi Yang}, \bibinfo{person}{Jiancan Wu},
  \bibinfo{person}{Zhicai Wang}, \bibinfo{person}{Xiang Wang},
  \bibinfo{person}{Yancheng Yuan}, {and} \bibinfo{person}{Xiangnan He}.}
  \bibinfo{year}{2023}\natexlab{}.
\newblock \showarticletitle{Generate what you prefer: Reshaping sequential
  recommendation via guided diffusion}.
\newblock \bibinfo{journal}{\emph{Advances in Neural Information Processing
  Systems}}  \bibinfo{volume}{36} (\bibinfo{year}{2023}),
  \bibinfo{pages}{24247--24261}.
\newblock


\bibitem[Zang et~al\mbox{.}(2022)]%
        {zang2022survey}
\bibfield{author}{\bibinfo{person}{Tianzi Zang}, \bibinfo{person}{Yanmin Zhu},
  \bibinfo{person}{Haobing Liu}, \bibinfo{person}{Ruohan Zhang}, {and}
  \bibinfo{person}{Jiadi Yu}.} \bibinfo{year}{2022}\natexlab{}.
\newblock \showarticletitle{A survey on cross-domain recommendation:
  taxonomies, methods, and future directions}.
\newblock \bibinfo{journal}{\emph{ACM Transactions on Information Systems}}
  \bibinfo{volume}{41}, \bibinfo{number}{2} (\bibinfo{year}{2022}),
  \bibinfo{pages}{1--39}.
\newblock


\bibitem[Zhao et~al\mbox{.}(2023)]%
        {zhao2023cross}
\bibfield{author}{\bibinfo{person}{Chuang Zhao}, \bibinfo{person}{Hongke Zhao},
  \bibinfo{person}{Ming He}, \bibinfo{person}{Jian Zhang}, {and}
  \bibinfo{person}{Jianping Fan}.} \bibinfo{year}{2023}\natexlab{}.
\newblock \showarticletitle{Cross-domain recommendation via user interest
  alignment}. In \bibinfo{booktitle}{\emph{Proceedings of the ACM web
  conference 2023}}. \bibinfo{pages}{887--896}.
\newblock


\bibitem[Zhou et~al\mbox{.}(2025)]%
        {zhou2025contrastive}
\bibfield{author}{\bibinfo{person}{Donglin Zhou}, \bibinfo{person}{Xinbei Cai},
  {and} \bibinfo{person}{Weike Pan}.} \bibinfo{year}{2025}\natexlab{}.
\newblock \showarticletitle{Contrastive Text-enhanced Transformer for
  Cross-Domain Sequential Recommendation}. In
  \bibinfo{booktitle}{\emph{Proceedings of the 31st ACM SIGKDD Conference on
  Knowledge Discovery and Data Mining V. 2}}. \bibinfo{pages}{4110--4119}.
\newblock


\bibitem[Zhu et~al\mbox{.}(2019)]%
        {zhu2019dtcdr}
\bibfield{author}{\bibinfo{person}{Feng Zhu}, \bibinfo{person}{Chaochao Chen},
  \bibinfo{person}{Yan Wang}, \bibinfo{person}{Guanfeng Liu}, {and}
  \bibinfo{person}{Xiaolin Zheng}.} \bibinfo{year}{2019}\natexlab{}.
\newblock \showarticletitle{Dtcdr: A framework for dual-target cross-domain
  recommendation}. In \bibinfo{booktitle}{\emph{Proceedings of the 28th ACM
  international conference on information and knowledge management}}.
  \bibinfo{pages}{1533--1542}.
\newblock


\bibitem[Zhu et~al\mbox{.}(2021)]%
        {zhu2021transfer}
\bibfield{author}{\bibinfo{person}{Yongchun Zhu}, \bibinfo{person}{Kaikai Ge},
  \bibinfo{person}{Fuzhen Zhuang}, \bibinfo{person}{Ruobing Xie},
  \bibinfo{person}{Dongbo Xi}, \bibinfo{person}{Xu Zhang},
  \bibinfo{person}{Leyu Lin}, {and} \bibinfo{person}{Qing He}.}
  \bibinfo{year}{2021}\natexlab{}.
\newblock \showarticletitle{Transfer-meta framework for cross-domain
  recommendation to cold-start users}. In \bibinfo{booktitle}{\emph{Proceedings
  of the 44th international ACM SIGIR conference on research and development in
  information retrieval}}. \bibinfo{pages}{1813--1817}.
\newblock


\bibitem[Zhu et~al\mbox{.}(2022)]%
        {zhu2022personalized}
\bibfield{author}{\bibinfo{person}{Yongchun Zhu}, \bibinfo{person}{Zhenwei
  Tang}, \bibinfo{person}{Yudan Liu}, \bibinfo{person}{Fuzhen Zhuang},
  \bibinfo{person}{Ruobing Xie}, \bibinfo{person}{Xu Zhang},
  \bibinfo{person}{Leyu Lin}, {and} \bibinfo{person}{Qing He}.}
  \bibinfo{year}{2022}\natexlab{}.
\newblock \showarticletitle{Personalized transfer of user preferences for
  cross-domain recommendation}. In \bibinfo{booktitle}{\emph{Proceedings of the
  fifteenth ACM international conference on web search and data mining}}.
  \bibinfo{pages}{1507--1515}.
\newblock


\end{thebibliography}

\end{document}